\documentclass[12pt]{article} 
\usepackage{epsfig,cite}
\usepackage{amsmath}
\usepackage{amssymb} 
\usepackage{color}
\usepackage{graphicx}
\usepackage{verbatim,mathrsfs,subfigure,feynmf}
\usepackage{url}
\usepackage{ytableau} 
\newcount\timecount
\newcount\hours \newcount\minutes  \newcount\temp \newcount\pmhours
\hours = \time
\divide\hours by 60
\temp = \hours
\multiply\temp by 60
\minutes = \time
\advance\minutes by -\temp
\def\hour{\the\hours}
\def\minute{\ifnum\minutes<10 0\the\minutes
            \else\the\minutes\fi}
\def\clock{
\ifnum\hours=0 12:\minute\ AM
\else\ifnum\hours<12 \hour:\minute\ AM
      \else\ifnum\hours=12 12:\minute\ PM
            \else\ifnum\hours>12
                 \pmhours=\hours
                 \advance\pmhours by -12
                 \the\pmhours:\minute\ PM
                 \fi
            \fi
      \fi
\fi
}

\def\monthname{\relax\ifcase\month 0/\or January\or February\or
   March\or April\or May\or June\or July\or August\or September\or
   October\or November\or December\else\number\month/\fi}

\def\bold#1{\setbox0=\hbox{$#1$}%
     \kern-.025em\copy0\kern-\wd0
     \kern.05em\copy0\kern-\wd0
     \kern-.025em\raise.0433em\box0 }

\def\beq{\begin{equation}}
\def\eeq{\end{equation}}

\def\ga{\mathrel{\raise.3ex\hbox{$>$\kern-.75em\lower1ex\hbox{$\sim$}}}}
\def\la{\mathrel{\raise.3ex\hbox{$<$\kern-.75em\lower1ex\hbox{$\sim$}}}}
\def\gev{{\rm \, Ge\kern-0.125em V}}
\def\tev{{\rm \, Te\kern-0.125em V}}
\def\gyr{{\rm \, G\kern-0.125em yr}}
\def\ohsq{\Omega_{\chi} h^2}
\def\slash#1{\rlap{\hbox{$\mskip 1 mu /$}}#1}%

\def\gappeq{\mathrel{\rlap {\raise.5ex\hbox{$>$}}
{\lower.5ex\hbox{$\sim$}}}}
\def\lappeq{\mathrel{\rlap{\raise.5ex\hbox{$<$}}
{\lower.5ex\hbox{$\sim$}}}}
\def\Toprel#1\over#2{\mathrel{\mathop{#2}\limits^{#1}}}

\def\m12{m_{1\!/2}}
\usepackage{slashbox}

\def\bea{\begin{eqnarray}}
\def\eea{\end{eqnarray}}

 \newcommand{\GeV}{\; \mathrm{GeV}}
 \newcommand{\TeV}{\; \mathrm{TeV}}
\def\gm2{g_{\mu}-2}

\newcommand{\xvev}[1]{\left\langle #1\right\rangle}  

\makeatletter 
\def\slash{\@ifnextchar[{\fmsl@sh}{\fmsl@sh[0mu]}} 
\def\fmsl@sh[#1]#2{% 
  \mathchoice 
    {\@fmsl@sh\displaystyle{#1}{#2}}% 
    {\@fmsl@sh\textstyle{#1}{#2}}% 
    {\@fmsl@sh\scriptstyle{#1}{#2}}% 
    {\@fmsl@sh\scriptscriptstyle{#1}{#2}}} 
\def\@fmsl@sh#1#2#3{\m@th\ooalign{$\hfil#1\mkern#2/\hfil$\crcr$#1#3$}} 
\makeatother 
\usepackage{slashbox}

\def\beq{\begin{equation}}
\def\eeq{\end{equation}}

\begin{document}
\begin{titlepage}
\pagestyle{empty}
\begin{flushright}
{\tt KCL-PH-TH/2024-41}, {\tt CERN-TH-2024-109}  \\
{\tt UMN-TH-4325/24, FTPI-MINN-24/16} \\
\end{flushright}

\vspace{0.5cm}
\begin{center}
{\bf \large{Non-universal SUSY models, $g_\mu-2$, $m_H$ and dark matter}}\\
\vskip 0.2in
{\bf John~Ellis}$^{a}$,
{\bf Keith~A.~Olive}$^{b}$ and
{\bf Vassilis~C.~Spanos}$^{b,c}$  
%}
\vskip 0.2in
{\small
{\em $^a$Theoretical Particle Physics and Cosmology Group, Department of
  Physics, King's~College~London, London WC2R 2LS, United Kingdom;\\
Theoretical Physics Department, CERN, CH-1211 Geneva 23,
  Switzerland}\\[0.2cm]
  {\em $^b$William I. Fine Theoretical Physics Institute, School of
 Physics and Astronomy,\\ University of Minnesota, Minneapolis, MN 55455,
 USA}\\[0.2cm]
{\em $^c$Section of Nuclear and Particle Physics, Department of Physics, \\
National and Kapodistrian University of Athens, 
   GR-157 84 Athens, Greece}\\[0.2cm] 
}

\vspace{1.5cm}
{\bf Abstract}
 \end{center}
\baselineskip=18pt \noindent
%%%%%%%%%%%%%%%%%%%%%%%%%%%%%%%%%%%%%%%%%%%%%%%%%
{\small
We study the anomalous magnetic moment of the muon, $g_\mu - 2 \equiv 2 a_\mu$, in the context of 
supersymmetric models  beyond the  CMSSM, where the  unification of either the gaugino masses $M_{1,2,3}$ or sfermion and Higgs masses is relaxed, taking into account the measured mass of the Higgs boson, $m_H$, the cosmological dark matter density and the direct detection rate.
We find that the model with non-unified  gaugino masses can make a contribution $\Delta a_\mu\sim 20 \times 10^{-10}$ to the anomalous magnetic moment of the muon, for example if 
$M_{1,2} \sim 600\GeV $ and  $M_3\sim 8 \TeV$. The model with non-universal sfermion and Higgs masses can provide even larger $\Delta a_\mu \sim 24 \times 10^{-10}$ if the sfermion masses for  the first and the second generations are $ \sim 400  \GeV$ and that of the 
third is $ \sim 8  \TeV$. 
We discuss the prospects for collider searches for supersymmetric particles in specific benchmark scenarios illustrating these possibilities, focusing in particular on the prospects for detecting the lighter smuon and the lightest neutralino.
}

%%%%%%%%%%%%%%%%%%%%%%%%%%%%%%%%%%%%%%%%%%%%%%%%

\vfill
%\leftline{CERN-PH-TH/2010-???}
\leftline{July 2024}

\end{titlepage}
%\baselineskip=18pt
%%%%%%%%%%%%%%%%%%%%%%%%%%%%%%%%%%%%%%%%%%%%%%%%%%

%%%%%%%%%%%%%%%%%%%%%%%%%%%%%%%%%%%%%%%%%%%%%%%%
\section{Introduction}
%%%%%%%%%%%%%%%%%%%%%%%%%%%%%%%%%%%%%%%%%%%%%%%%
\label{sec:intro}
It has been more than two decades since the initial report  of a deviation, $\Delta a_\mu$   between 
the experimental measurement of the muon anomalous magnetic moment  $g_\mu - 2 \equiv 2 a_\mu$ 
and the Standard Model theoretical prediction~\cite{BNL1}. The significance of this ``anomalous anomaly" grew 
over time with the enhanced precision of the BNL experiments~\cite{BNL2} and the 
recent measurements from the Fermilab experiment~\cite{FNAL}. 
On the theory side, 
the data-driven Standard Model theoretical predictions were refined,
thanks to better estimations of the hadronic vacuum polarization and light-by-light
scattering effects~\cite{Theory}.
The combined BNL and Fermilab results yield~\cite{Muong-2:2023cdq} 
\beq
a_\mu^{exp}=  116 592 059(22)   \times 10^{-11}
\eeq   
whereas the data-driven theoretical value is~\cite{Theory}
\beq
a_\mu^{th}= 116 591 810(43)  \times 10^{-11} \, ,
\eeq   
corresponding to a discrepancy of $\Delta a_\mu = (24.9 \pm 4.8) \times 10^{-10} $. 

However, the uncertainties in first-principles lattice calculations have now been reduced to a level
comparable to the data-driven calculation. Moreover, 
a pioneering lattice calculation of the intermediate-scale ``window function"~\cite{Lattice},
when extrapolated to larger and smaller scales, 
yielded a central value of $a_\mu$ that is in significant tension with
the data-driven estimate and corresponds to a smaller  discrepancy
of $ \Delta a_\mu = (10.7 \pm 6.9) \times 10^{-10}$. Subsequently, several other lattice calculations \cite{Kuberski:2023qgx}
of the window function have yielded numerical results that are similar to those of~\cite{Lattice}.

More recently, the CMD-3 Collaboration has published new $e^+ e^- \to 2 \pi$ production data \cite{cmd} in the
energy range $E_{CM} < 1$~GeV that are in significant tension with the previous world average results
used in~\cite{Theory}, and correspond to an even smaller discrepancy
of $\Delta a_\mu = (4.9 \pm 5.5) \times 10^{-10}$. This has led a reappraisal of previous
results on $e^+ e^- \to 2 \pi$ production and of estimates of the low-energy vacuum
polarization using $\tau \to \nu +$~hadrons data, leading to an estimated discrepancy
of $\Delta a_\mu = (12.3 \pm 4.9) \times 10^{-10}$ \cite{Davier:2023fpl}.

\begin{figure}[ht!]
\centering
\includegraphics[width=0.75\textwidth]{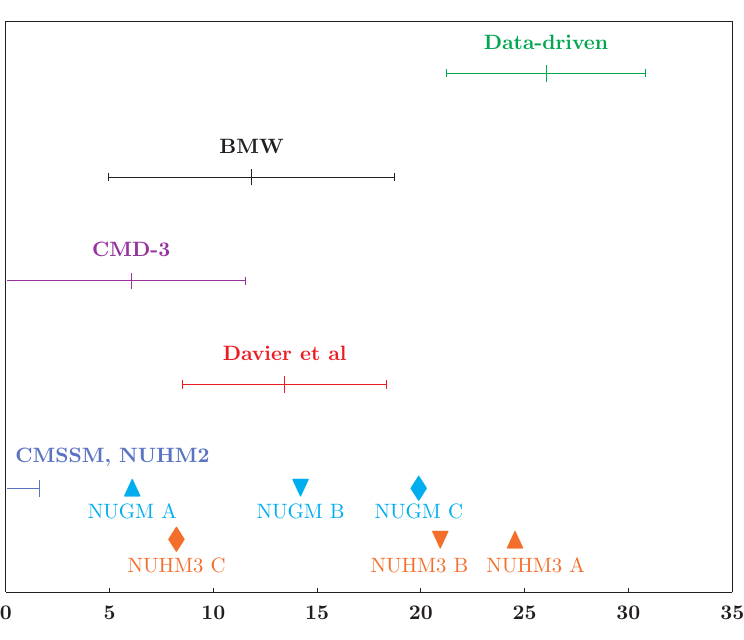}
\caption{Comparison of experimental and theoretical estimates of
$\Delta a_\mu$~\cite{Theory,Lattice,cmd,Davier:2023fpl} with calculations in supersymmetric models
including the benchmarks introduced in this paper.}
\label{fig:g-2_Summary_Table}
\end{figure}

Figure~\ref{fig:g-2_Summary_Table} displays the estimates of $\Delta a_\mu$ introduced above:
the data-driven estimate~\cite{Theory}, the lattice calculation in~\cite{Lattice}, 
the estimate based on the recent CMD-3 experimental measurement~\cite{cmd}, and the experimental
re-evaluation in~\cite{Davier:2023fpl}.

Following the initial BNL experimental result, theories based on supersymmetry were quickly 
suggested as explanations for the apparent discrepancy between experiment and the data-driven
theoretical calculations within the Standard Model~\cite{ENO,g-2}. These calculations
motivated light smuon and neutralino masses of a few hundred GeV. 
However, the enthusiasm for supersymmetric theories has diminished
subsequently, as direct evidence for supersymmetry has remained elusive,
particularly in the Large Hadron Collider (LHC) experiments~\cite{LHCSUSY}. When the
LHC results are interpreted within the Constrained Minimal Supersymmetric extension of the
Standard Model (CMSSM), in which all the gaugino masses and all the sfermion masses are assumed 
to be universal at some high unification scale, the possible supersymmetric contribution
to $a_\mu$ is constrained to be much smaller \cite{otherCMSSM,Ellis:2021zmg,Wang:2021bcx} than the discrepancy between the experimental data 
and the data-driven theoretical estimate recommended in ~\cite{Theory}, as seen in Fig.~\ref{fig:g-2_Summary_Table}.

However, for several reasons the comparison between supersymmetric models and the data merits a more thorough study.
First, the new lattice calculations and data on the hadron vacuum polarization
suggest that the discrepancy between experimental data and the Standard Model may be smaller
than thought previously. Secondly, there are many possible generalizations of the
CMSSM that could accommodate a light smuon and a light neutralino mass simultaneously with the heavier
squarks and gluinos indicated by the unsuccessful LHC searches. Thirdly, 
the lower limits on sparticle masses set in simplified models often do not 
apply to more general models.

We show in this paper that a significant supersymmetric contribution to $a_\mu$ that could
match any of the estimates of the discrepancy between the Standard Model and the BNL and Fermilab
measurements mentioned above, i.e., $\Delta a_\mu \sim 250, 100$ or 50 if the specific
unification conditions on the gaugino and/or sfermion masses are relaxed.
We illustrate this point within two specific generalizations of the CMSSM, one in which
the gaugino masses $M_{1,2,3}$ are non-universal \cite{Wang:2018vrr,Wang:2021bcx,Aboubrahim:2021xfi} (the NUGM), and one in which unification of the masses 
of the first- and second-generation sfermions, the third-generation sfermions and the Higgs 
scalar multiplets is relaxed \cite{Baer:2021aax} (the NUHM3). In both cases, we perform an extensive scan of the parameter space 
using Markov Chain Monte Carlo (MCMC) techniques similar to those used previously to scan the CMSSM parameter space \cite{Ellis:2022emx}, which are geared to providing a Higgs mass and relic density for the lightest supersymmetric particle (LSP) compatible with experimental constraints~\cite{Aad:2012tfa,Planck}.
We also highlight benchmark scenarios that illustrate
the phenomenological possibilities, taking into account also calculations of the 
mass of the Higgs boson, $m_H$, and constraints on the density of relic cold dark matter and
its possible spin-independent and -dependent cross-sections for scattering on matter.

The values of $\Delta a_\mu$ attainable in the NUGM and NUHM3 are illustrated by these benchmark scenarios, as indicated in Fig.~\ref{fig:g-2_Summary_Table}.
As shown there, we find that a model with non-universal  gaugino masses (the NUGM) can yield $\Delta a_\mu\sim 20 \times 10^{-10}$, sufficient to accommodate the lattice and CMD-3 results, e.g., if 
$M_{1,2} \sim 600\GeV $ and  $M_3\sim 8 \TeV$. In the case of non-universal sfermion and Higgs masses (the NUHM3),
we find that even larger $\Delta a_\mu \sim 24 \times 10^{-10}$ can be reached, sufficient to accommodate the data-driven value given in~\cite{Theory}, for first- and second-generation sfermion masses $ \sim 400  \GeV$ and third-generation sfermion masses  $ \sim 8  \TeV$. The Planck~\cite{Planck}
value of the cosmological dark matter density can be saturated by the LSP density in the NUGM, but
not in the NUHM3 because the spin-independent dark matter scattering cross section is larger in that
model.

The paper is organized as follows. In Section~\ref{sec:models} we introduce the NUGM and NUHM3 models that we study, and in Section~\ref{sec:proc} we describe our analysis procedure. Our results are presented in Section~\ref{sec:results}: those for the NUGM in Subsection~\ref{sec:NUGM} and those for the NUHM3 in Subsection~\ref{sec:NUHM3}. Section~\ref{sec:benchmarks} focuses on some benchmark scenarios that illustrate the phenomenological possibilities in these models, including the possibilities for probing them at the LHC and elsewhere, as well as their possible contributions to $a_\mu$.
Finally, Section~\ref{sec:concl} summarizes our conclusions.

%%%%%%%%%%%%%%%%%%%%%%%%%%%%%%%%%%%%%%%%%%%%%%%%
\section{The Models}
%%%%%%%%%%%%%%%%%%%%%%%%%%%%%%%%%%%%%%%%%%%%%%%%
\label{sec:models}
As already mentioned, it is well known that a substantial discrepancy between the Standard Model and experimental measurements of the anomalous magnetic moment  of the muon $g_\mu -2$ can no longer be explained within the CMSSM \cite{otherCMSSM,Ellis:2021zmg,Wang:2021bcx,Aboubrahim:2021xfi,Baer:2021aax}. 
The models considered here are based on generalizations of the CMSSM \cite{DN,cmssm, eoss,interplay,Ellis:2015rya,Ellis:2018jyl,Ellis:2019fwf}, which is defined by four continuous parameters: a universal gaugino mass, $M_{1/2}$, a universal scalar mass, $m_0$, a universal trilinear mass, $A_0$ (all three of which are supersymmetry-breaking mass terms) and the ratio of the 
vacuum expectation value (vev) of the neutral components of the Higgs doublets
$\xvev{H_2^0}/\xvev{H_1^0}=\tan\beta$. In addition there is the freedom to choose the sign of the $\mu$-term. 
The leading supersymmetric contributions to $g_\mu -2$ arise from two one-loop graphs: one  involving  charginos and sneutrinos and another that involves smuons and neutralinos.~\footnote{See~\cite{Heinemeyer:2003dq} for the two-loop corrections, which are subdominant.} Our calculations of the contributions to $g_\mu - 2$ are detailed in \cite{ENO} and follow the calculations in \cite{IN}.
Substantial contributions to $g_\mu -2$ may arise if either the chargino/neutralino pair and/or the
smuon/neutralino pair weigh a few hundred GeV. However, in the CMSSM the Higgs mass $m_H=125 \GeV$ in conjunction with the constraints on direct detection of astrophysical dark matter put severe lower limits on both
the assumed universal  gaugino mass $M_{1/2}$ and the common sfermion mass $m_0$,
implying $m_0 \ge 10 \TeV$ and  $M_{1/2} \ge 4-5  \TeV$~\cite{Ellis:2022emx}. These limits 
imply that in the CMSSM all sleptons and gauginos are heavy, and consequently the supersymmetric contributions to $g_\mu-2$ are small, as we review below.

However, if one relaxes either the gaugino~\cite{Wang:2018vrr,Wang:2021bcx,Aboubrahim:2021xfi} or the sfermion~\cite{Baer:2021aax}
mass unification conditions, one may expect to find enhanced supersymmetric corrections to 
$g_\mu -2$. With this in mind,  we  explore two extensions of the CMSSM in which  
these mass unification conditions are relaxed:
\begin{itemize}
\item
{\bf Non-Universal gaugino masses}: A model in which
gaugino mass unification is abandoned (NUGM) \cite{nugm,Wang:2018vrr,Wang:2021bcx,Aboubrahim:2021xfi}, which has six free parameters, namely separate gaugino masses $M_1$, $M_2$, $M_3$ in addition to the other
CMSSM parameters, $m_0$, $A_0$, $\tan\beta$ and  $sgn(\mu)$. 

\item 
{\bf Non-universal sfermion and Higgs masses}: 
Two well studied extensions of the CMSSM allow for either one \cite{nuhm1} or both \cite{nuhm2} of the Higgs soft mass parameters to differ from the universal scalar masses.
These are known as the NUHM1 and NUHM2, respectively, with $m_{H_1} = m_{H_2} \ne m_0$ or $m_{H_1} \ne m_{H_2} \ne m_0$. The extension we study here (dubbed NUHM3 \cite{Baer:2021aax}) is
a model in which common masses of the first and second sfermion generations, $m_{012}$, differ from that
of the third, $m_{03}$, and those of the two Higgs multiplets, $m_{H_1}$ and   $m_{H_2}$, at the GUT scale.
It is convenient to use the minimization conditions of the effective electroweak potential
to trade $m_{H_1,H_2}$ for the Higgsino mixing  parameter $\mu$ and the mass of the pseudoscalar 
Higgs boson $A$, $M_A$.
The replacement of $m_{H_1,H_2}$ by  $\mu, M_A$ is the same as in the non-universal Higgs model (NUHM2) or
a general two Higgs doublet model (2HDM).   
This model is characterized by seven parameters:  $M_{1/2}$, $m_{012}$, $m_{03}$, $\mu$, $M_A$, 
$A_0$ and  $\tan\beta$. Since the sign of the Higgsino mixing parameter $\mu$ is the same  as  
that of the   supersymmetric contribution to  $g_\mu - 2$, we consider only positive values for this parameter. 
\end{itemize} 
A related study \cite{Ajaib:2023jhc} was performed allowing for non-universal gaugino masses and non-universal Higgs masses, with eight free continuous parameters. Models attempting to account for $\Delta a_\mu$ for which the relic density is determined by stau coannihilation \cite{stauco} were considered in \cite{sven} or by a well-tempered neutralino \cite{wt} in \cite{sven2}.
Even more generalized models form the basis of the pMSSM \cite{pMSSM0,pmssm}.

%%%%%%%%%%%%%%%%%%%%%%%%%%%%%%%%%%%%%%%%%%%%%%%%
\section{Analysis Procedure}
%%%%%%%%%%%%%%%%%%%%%%%%%%%%%%%%%%%%%%%%%%%%%%%%
\label{sec:proc}

We have scanned the six-dimensional parameter space of the NUGM and the
seven-dimensional parameter space of the NUHM3 with MCMC routines 
using initial likelihood functions for the following three basic predictions of the model: the 
Higgs mass, $m_H$, the neutralino relic density, $\ohsq$, and the muon dipole moment, $g_\mu-2$. 
The likelihood functions are fixed by the following experimental ranges: $m_H=125 \pm 2 \GeV$,
$\ohsq=0.12 \pm 0.0012$ and  $\Delta a_\mu = (24.9 \pm 4.8) \times  10^{-10}$.  
These ranges are chosen to aid the parameter scan, but do not directly affect our results. 
The range of $m_H$ is broader than the current estimated uncertainty in MSSM calculations, 
and is much larger than the experimental uncertainty. However, it is consistent at 2 $\sigma$ with the theoretical uncertainty in the calculation of $m_H$ using {\tt FeynHiggs} \cite{FH}, which we adopt in all the results below. We present results assuming that $\ohsq$ either provides all of the cosmological cold dark matter density~\cite{Planck},
or that there is potentially another contribution to the cold dark matter density. Likewise, motivated by the recent lattice and experimental results,
we consider the possibility that $\Delta a_\mu$ is smaller than the range quoted above.

We assume that the lightest neutralino is the lightest supersymmetric particle (LSP), and that $R$-parity is conserved, so that the LSP is stable.
We apply the same accelerator constraints as in \cite{Ellis:2021zmg}, namely:

\begin{itemize}

\item The LEP experiments' exclusion of charginos lighter than 103 GeV 
 if  $ m_{\tilde{\chi}^{\pm}}- m_\chi \geq 3\,\GeV$~\cite{LEPSUSY1} and
$m_{\tilde{\chi}^{\pm}} > 91.9\ \GeV$ for $m_{\tilde{\chi}^{\pm}}-  m_\chi < 3\,\GeV$~\cite{LEPSUSY2}.

    \item In both the models we study, the smuon and selectron masses are equal up to small
    corrections related to the muon-electron mass difference. The LEP experiments established lower limits on selectron masses that are generally stronger than those on smuons, and stronger for left-handed sleptons. For our purposes the most relevant LEP slepton constraints are those on
    $m_{\tilde e_R}$, which also depend on other sparticle masses, in particular $m_\chi$~\cite{PDG}. 
    We use a LEP lower limit of $100$~GeV in general, reducing to 73~GeV if  
    $m_{\tilde \mu_R} - m_\chi \lesssim 2$~GeV.

\item At the LHC, ATLAS has established the lower limit $m_{\tilde \mu} \gtrsim 700$~GeV
when $m_\chi = 0$,
falling to $\gtrsim 600$~GeV when $m_\chi \simeq 400$~GeV~\cite{slepton}.
At lower smuon masses there is an allowed corridor where $m_{\tilde \mu} - m_\chi \lesssim 100$~GeV that extends down to the LEP lower limit on $m_{\tilde \mu}$. We have implemented fully these LHC limits in our scan of the sparticle parameter space: see~\cite{ATLAS:2023xco} for a full description.

\item An additional LHC constraint is relevant for compressed spectra when $m_{\mu_R} - m_\chi 
\lesssim 15$~GeV~\cite{Aad:2019qnd}, which is maximized when 
$m_{\tilde \mu_R} - m_\chi \simeq 10$~GeV, in which  case it excludes 
$m_{\tilde \mu_R} \lesssim 150$~GeV. 

\item We do not impose {\it a priori} LHC constraints on heavier sparticles that are, in general, more model-dependent. (See, in particular, the second paper in~\cite{pmssm}.) However, we do comment {\it a posteriori} on their potential significances for benchmark NUGM and NUHM3 scenarios.

\item As noted earlier, the Higgs mass is constrained to lie in the range $123~{\rm GeV} \le m_H \le 127~{\rm GeV}$ as computed using {\tt FeynHiggs} \cite{FH}.

\item For cases where the LSP makes up all of the cold dark matter density, we require $\Omega_\chi h^2 = 0.1200 \pm 0.0036$ which represents the 3$\sigma$ range as determined by Planck \cite{Planck}. When the LSP is allowed to make up only a fraction of the dark matter, we simply exclude models which produce $\Omega h^2 > 0.1236$. 

\end{itemize}

We also compute in our analysis the spin-dependent and -independent cross sections for the scattering of the LSP on protons. For a description of our calculations of these, see \cite{sospin}. The resulting cross sections are compared to the recent results from LZ \cite{LZ} for the spin-independent scattering cross section and PICO \cite{Pico} for the spin-dependent scattering cross section on protons.
Note that while we exclude input parameters which yield a relic density in excess of the Planck determination, we do not exclude {\it a priori} points that yield a scattering cross section in excess of the current experimental limits, but comment {\it a posteriori} on their impacts on the NUGM and NUHM3 parameter spaces.

%%%%%%%%%%%%%%%%%%%%%%%%%%%%%%%%%%%%%%%%%%%%%%%%
\section{Results}
%%%%%%%%%%%%%%%%%%%%%%%%%%%%%%%%%%%%%%%%%%%%%%%%
\label{sec:results}

\subsection{Non-Universal Gaugino Mass Model (NUGM)}
\label{sec:NUGM}

As described above, the NUGM is characterized by six parameters. 
In all cases, we have chosen $\mu > 0$. In our scan of the parameter space, we have covered the following ranges 
\begin{align}
M_1 & = 0 - 10 \TeV \, , \nonumber \\
M_2 & =  0 - 10 \TeV \, , \nonumber \\
M_3 & =  0 - 15  \TeV \, , \nonumber \\
m_0 & = 0 - 10 \TeV   \, , \nonumber \\
|A_0/m_0| & = 0 - 40 \, , \nonumber \\
\tan \beta & = 1 - 50 \, .
\label{r1}
\end{align}
The supersymmetry-breaking mass parameters are input at the GUT scale, taken to be where the two electroweak gauge couplings are unified, and run down to the electroweak scale. Any set of inputs that do not yield a neutralino LSP is discarded. In addition, we discard parameter choices that lead to a tachyonic Higgs pseudoscalar or do not satisfy the electroweak symmetry breaking minimization conditions, i.e., that lead to $\mu^2 < 0$.~\footnote{We also made an exploratory scan with negative values of $M_i$. Changing the sign of $M_1$
flips the sign of $\Delta a_\mu$, which can be compensated by choosing $\mu < 0$. We checked that there were no qualitative effects on our results for $M_{2,3}$. Therefore in all that follows, we choose to keep $M_i > 0$ and $\mu > 0$.}

In our scan of the six-dimensional parameter space of the NUGM  we obtain   250,000 points from the MCMC code. About  60,000 of them pass the kinematical constraints described in Section~\ref{sec:proc} and   satisfy the requirement that 
LSP is the lightest neutralino. Of these,  45,000 points  satisfy the Higgs boson mass bound    
$m_H=125 \pm 2 \GeV$ and the relic density requirement   $\Omega_\chi h^2 \leq 0.1236$. 
Of these,  3,000 points yield an LSP density within the Planck $3\sigma $ range for the overall cold dark matter density:  $0.1164 \leq \Omega_\chi h^2 \leq 0.1236$.

\begin{figure}[ht!]
\centering
\includegraphics[width=0.45\textwidth]{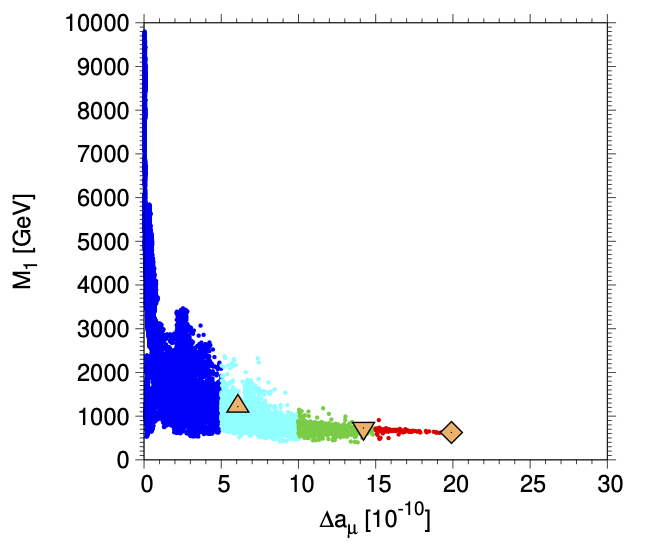}
 \includegraphics[width=0.45\textwidth]{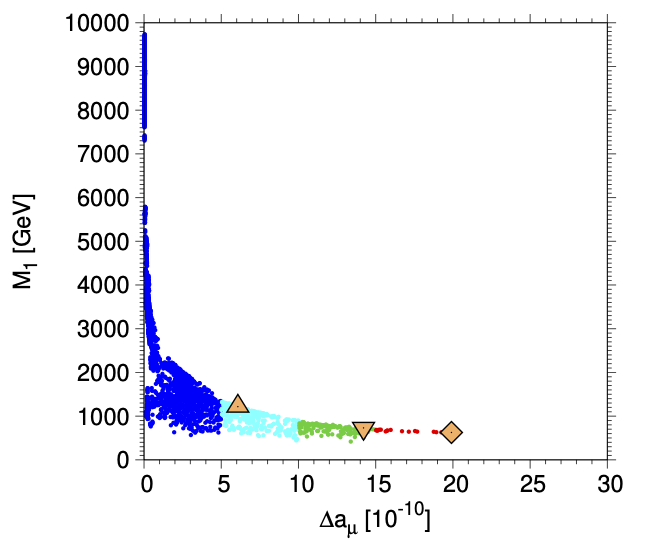}\\
 \vspace{-2mm}
 \includegraphics[width=0.45\textwidth]{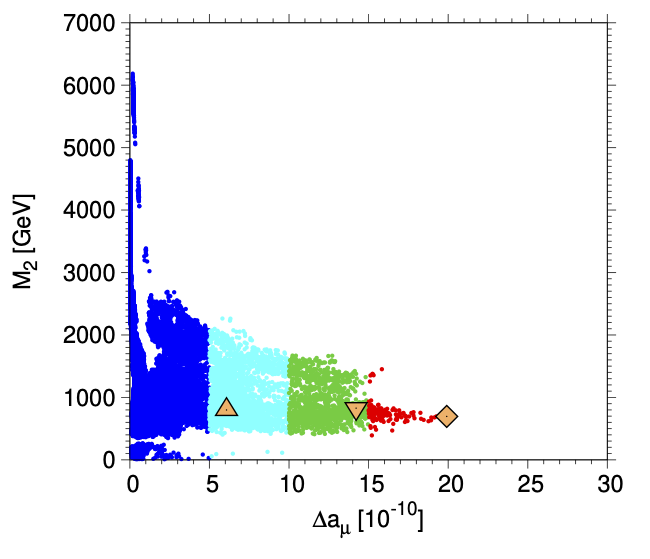}
 \includegraphics[width=0.45\textwidth]{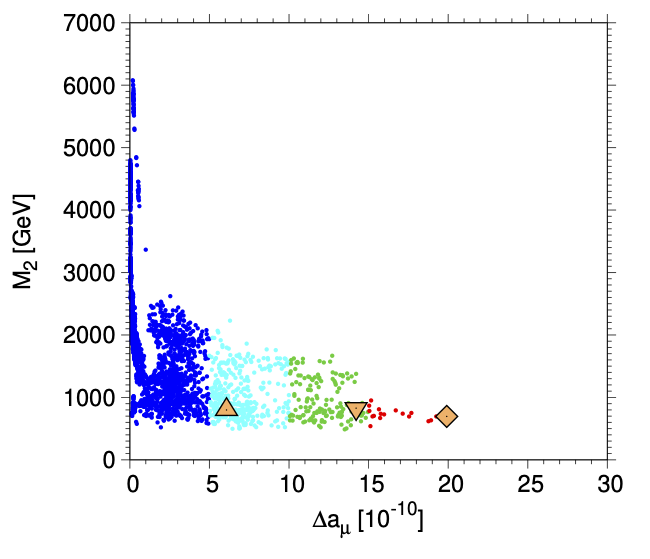}\\
 \vspace{-2mm}
\includegraphics[width=0.45\textwidth]{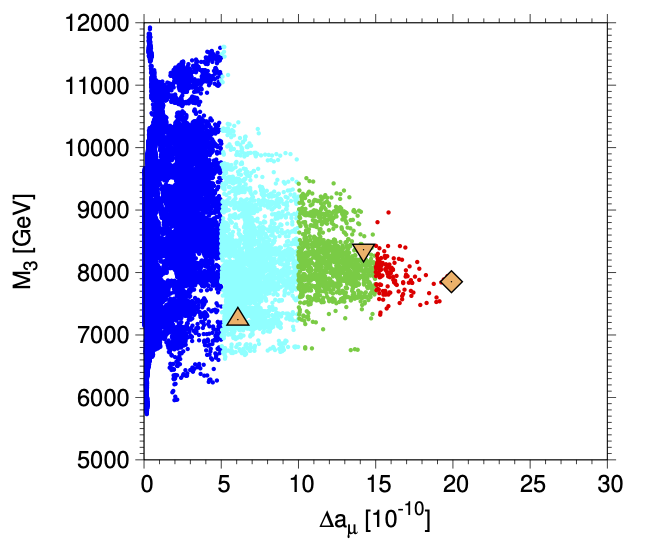}
 \includegraphics[width=0.45\textwidth]{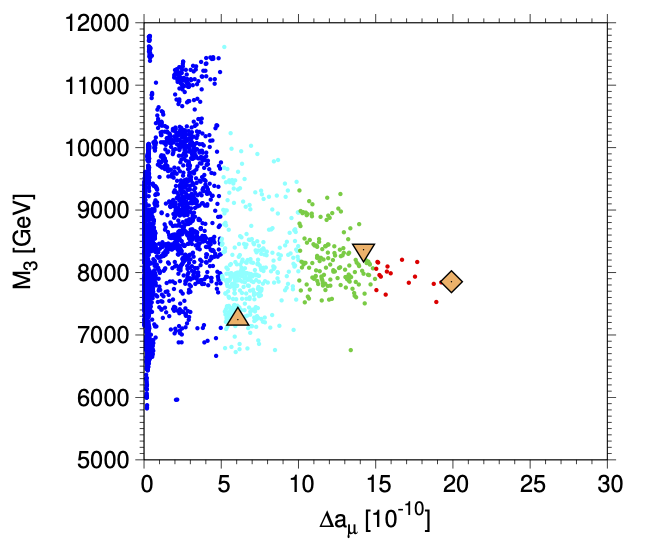}\\
\caption{\it  Allowed values of
the gaugino masses $M_{1,2,3}$ in the NUGM as functions of $\Delta a_\mu$. The left panels are for 
$\Omega_\chi h^2 \le 0.1236$, whereas the right panels are for points
with $0.1164 \le \Omega_\chi h^2 \le 0.1236$. {The yellow symbols indicate the locations of the NUGM benchmark points discussed in Section~\ref{sec:benchmarks} below.}
}
\label{fig:Gaugino_masses}
\end{figure}

\begin{figure}[ht!]
\centering
\includegraphics[width=0.45\textwidth]{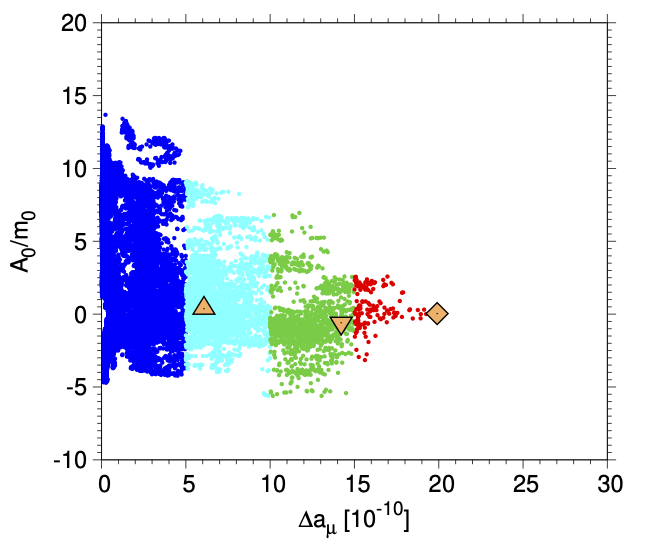}
 \includegraphics[width=0.45\textwidth]{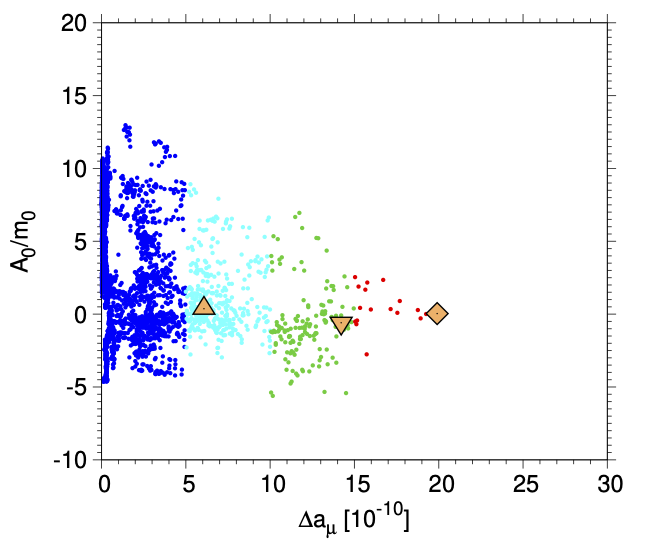}\\
 \vspace{-2mm}
 \includegraphics[width=0.45\textwidth]{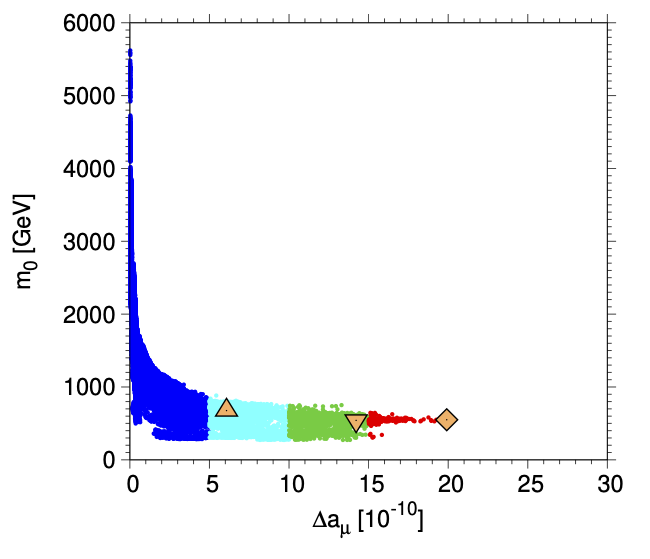}
 \includegraphics[width=0.45\textwidth]{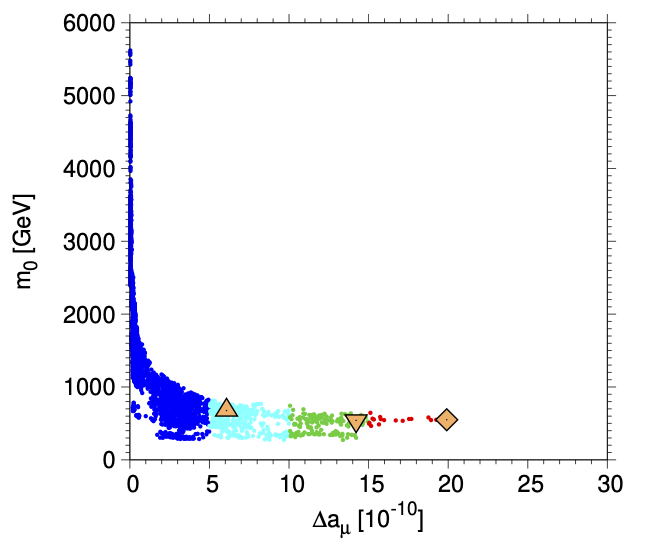}\\
 \vspace{-2mm}
\includegraphics[width=0.45\textwidth]{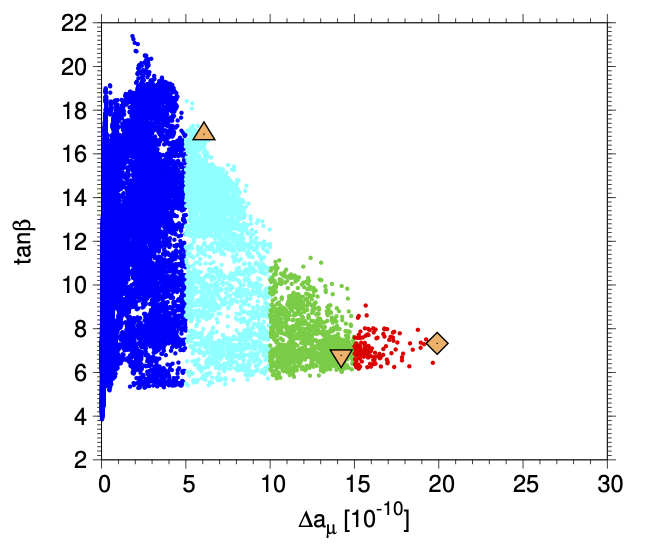}
 \includegraphics[width=0.45\textwidth]{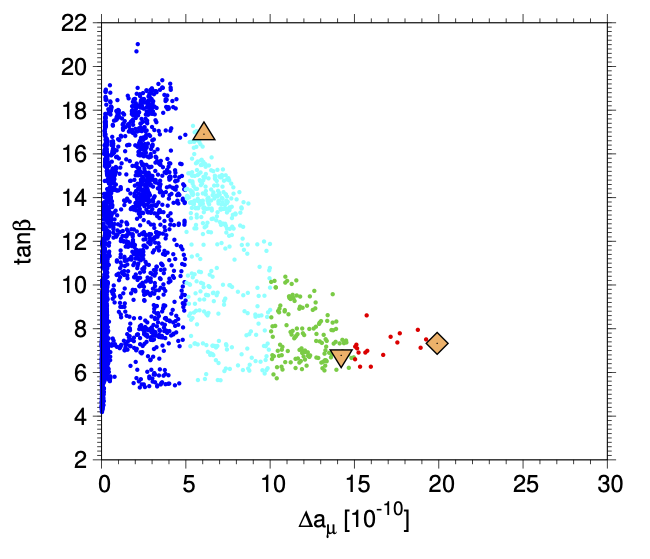}\\
\caption{\it As in Fig.~\ref{fig:Gaugino_masses}, showing the allowed values of
$A_0/m_0$, $m_0$ and $\tan\beta$ in the NUGM as functions of $\Delta a_\mu$.}
\label{fig:A0_m0_tb}
\end{figure}

%\vspace{2mm}
In Figs.~\ref{fig:Gaugino_masses} and \ref{fig:A0_m0_tb}
we show the values of the input parameters for points that survive all of the constraints discussed in the previous Section. The values are displayed as functions of the calculated value of $\Delta a_\mu$ for each point and color-coded in intervals of $\Delta a_\mu = 5 \times 10^{-10}$. Plots on the left use only the upper bound on $\Omega_\chi h^2$, whereas those on the right are restricted to the Planck determination of the cold dark matter density,
as indicated in the caption. As might be expected, we see in Fig.~\ref{fig:Gaugino_masses} that the viable resulting parameter sets indicate that $M_1 \sim M_2 \lesssim 1$~TeV for points with the largest contribution to $\Delta a_\mu \sim 20 \times 10^{-10}$.
However, the gluino mass $M_3 \sim 8$~TeV is significantly larger, as is required in order to increase the stop mass sufficiently to obtain a Higgs mass consistent with experiment. {The yellow symbols in these and subsequent planes indicate the locations of the NUGM benchmark points discussed in Section~\ref{sec:benchmarks} below.}

We see in the top panels of Fig.~\ref{fig:A0_m0_tb} that 
the least well determined input parameter is $A_0/m_0$, for which no specific value is indicated.  Indeed, 
$A_0/m_0$ is only very weakly
constrained for points that yield small values of $\Delta a_\mu \lesssim 5 \times 10^{-10}$, 
whereas relatively small values of $|A_0/m_0| \lesssim 2$ are favored for points 
that yield larger values of $\Delta a_\mu \gtrsim 15 \times 10^{-10}$. The middle panels of
Fig.~\ref{fig:A0_m0_tb} show, unsurprisingly, that larger values of $\Delta a_\mu$ are
correlated with smaller values of $m_0$. 
This preference for $m_0 \sim 0.5$~TeV
enables a suitably low value of $m_{\tilde \mu}$ to be obtained.
The bottom panels of Fig.~\ref{fig:A0_m0_tb}
show that, whereas small values of $\Delta a_\mu \lesssim 5 \times 10^{-10}$ are possible
for values of $\tan \beta \in [4, 20]$, large values of 
$\Delta a_\mu \gtrsim 15 \times 10^{-10}$ are found only for $\tan \beta \sim 8$.

To understand this, we recall that in order to get a sizeable  contribution to  $\Delta a_\mu$, a light smuon and LSP are preferred, so  a typical acceptable NUGM point has small $M_1$, $M_2$ and $m_0$. On the other 
hand, high values of $M_3$ are needed in order to obtain heavy stops that yield a Higgs mass in the required region $125 \pm 2 \GeV$. 
One then finds that the two-loop RGEs affect $m_{H_{1,2}}$ and increase $\mu$ which drives $m_{\tilde {\mu}_L}$ and   $m_{\tilde {\tau}_1}$ lighter for higher $M_3$,
favoring the stau coannihilation region, where $\Omega_\chi h^2 \simeq 0.12$. At $M_3 \gtrsim 8$~TeV, the stau becomes the LSP and ultimately tachyonic.  Similarly,  $\tan\beta$ must be $ \lesssim 10$, since higher values also result in a stau LSP.

\begin{figure}[ht!]
\centering
\includegraphics[width=0.45\textwidth]{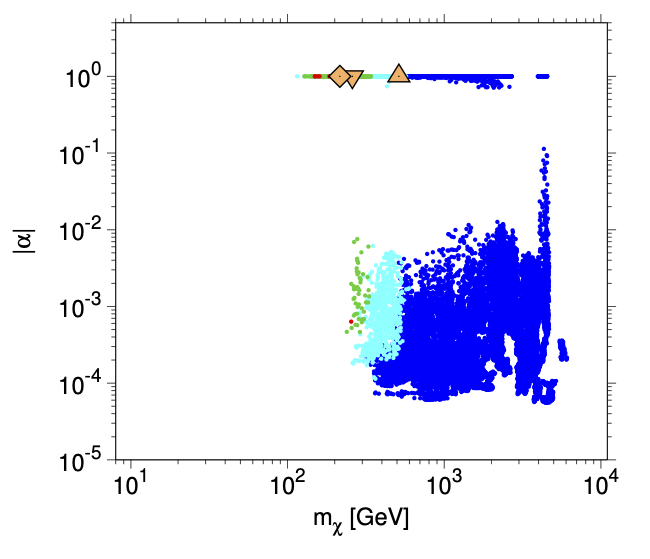}
 \includegraphics[width=0.45\textwidth]{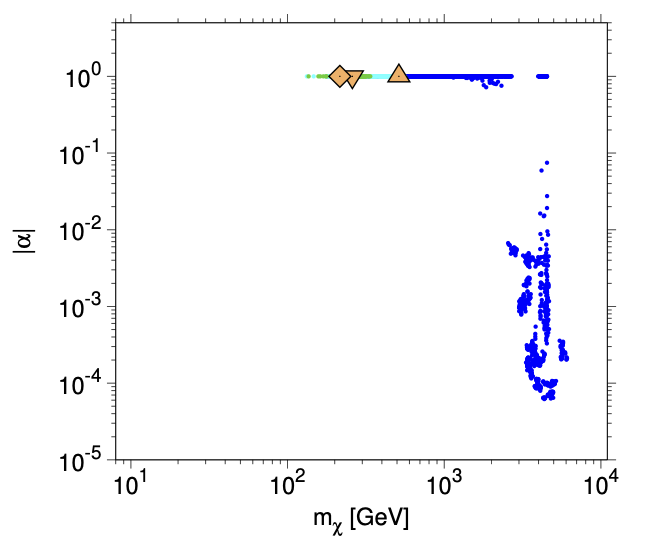}\\
  \vspace{-2mm}
\includegraphics[width=0.45\textwidth]{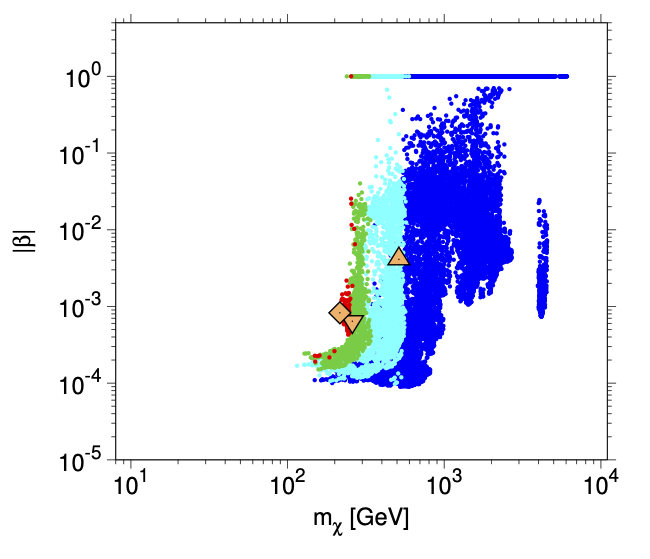}
 \includegraphics[width=0.45\textwidth]{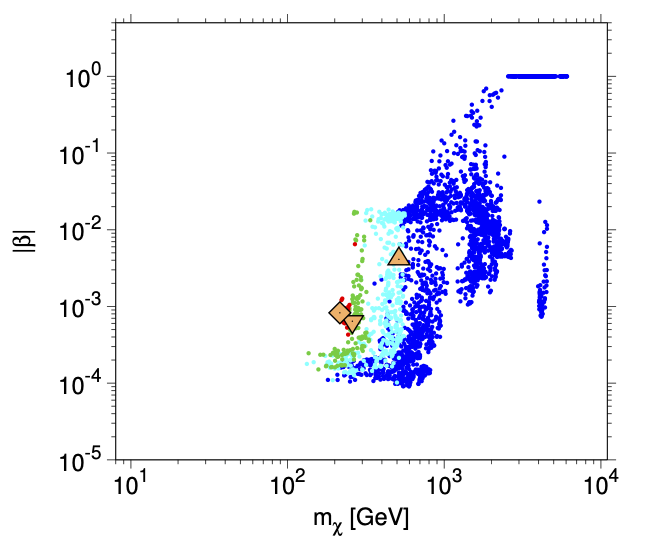}\\
 \caption{\it The Bino (upper panels) and Wino  (lower panels) components $\alpha$ and $\beta$, respectively, of the LSP for the NUGM points shown in Fig.~\ref{fig:Gaugino_masses} as functions of  $m_\chi$. The color coding corresponds to the values of $\Delta a_\mu$
 shown in Fig.~\ref{fig:Gaugino_masses}.
}
\label{fig:neutralino_content}
\end{figure}

The composition of the neutralino LSP, $\chi$, can  be expressed as a linear combination of the Bino (U(1) gaugino), neutral Wino (SU(2) gaugino), and two neutral Higgsinos
\begin{equation}
	\chi = \alpha \, \tilde B + \beta \, \tilde  W^3 + \gamma \, \tilde H^0_1 +
\delta \, \tilde H^0_2 \, ,
\label{chimix}
\end{equation}
whose masses and hence the neutralino composition are determined by the gaugino masses, $M_{1,2}$, the $\mu$ parameter and  $\tan \beta$. 
Figure~\ref{fig:neutralino_content} displays the gaugino contents, $\alpha$ and $\beta$, of the NUGM sample points.
We see that they mainly exhibit two possibilities: an almost pure Bino with a small admixture of the neutral Wino, or vice versa.
The points are color-coded according to their values of $\Delta a_\mu$ as in Fig.~\ref{fig:Gaugino_masses}.
We note that the majority of
the points with relatively large $\Delta a_\mu \gtrsim 15 \times 10^{-10}$ have a dominant
Bino component.
The most significant difference between the left panels 
(where the Planck range of the cold dark matter density is interpreted as an upper limit on the LSP density)
and the right panels (where the LSP is assumed to dominate the Planck density range)
is that in the latter case there is limited overlap between the ranges of $m_\chi$ where the
Bino and Wino components may dominate. 

For models in which the Bino is the LSP, its
relic density is determined by various annihilation and coannihilation channels. For relatively low Bino masses, annihilations in what was termed the bulk region of the the MSSM parameter space together with Bino-stau coannihilations dominate \cite{eoss}. In the CMSSM, these have long been excluded (at the time, by the LEP lower bound on the Higgs mass), but this is not an issue in the NUGM. At larger masses, Bino-Wino and Bino-chargino coannihilations \cite{chaco,esug} become important. When the LSP is predominantly a Wino, it is interesting to note that when the Sommerfeld enhancement \cite{Sommerfeld1931} of Wino annihilations is included, typically a narrow range of Wino masses with $m_{\tilde W} \simeq 3$~TeV  is required to attain $\Omega_{\tilde W} h^2 = 0.12$ \cite{winomass1,winomass,mc13}. This accounts for the large-mass end of the horizontal strip seen in the lower panels of Fig.~\ref{fig:neutralino_content}. The Wino relic density is typically smaller for lower masses, accounting for the elongated strip in the lower left panel. In the lower right panel the strip extends down to roughly 2 TeV, where nominally the relic density would be relatively low. However for these points, the relic density is controlled by Wino-Bino coannihilation, as the Bino is nearly degenerate with the Wino and both states contribute to the relic density, which is actually enhanced.

\begin{figure}[ht!]
\centering
\includegraphics[width=0.45\textwidth]{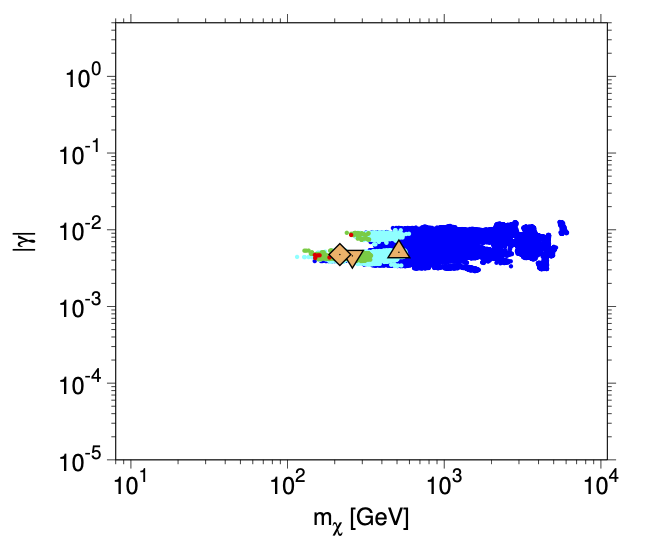} 
\includegraphics[width=0.45\textwidth]{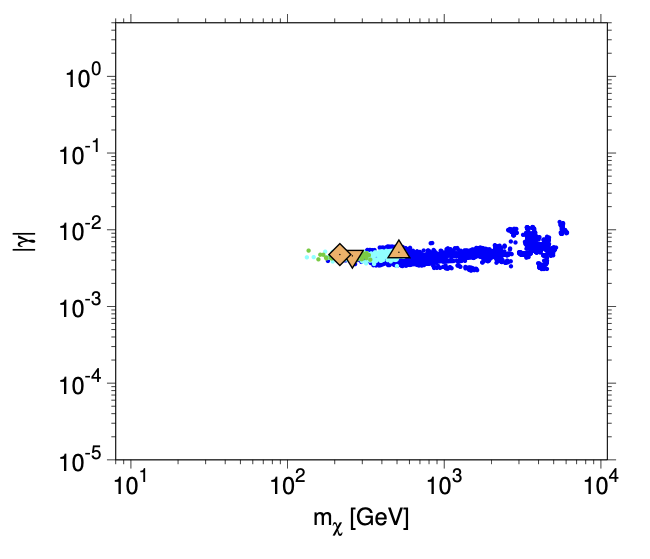}\\
  \vspace{-2mm}
\includegraphics[width=0.45\textwidth]{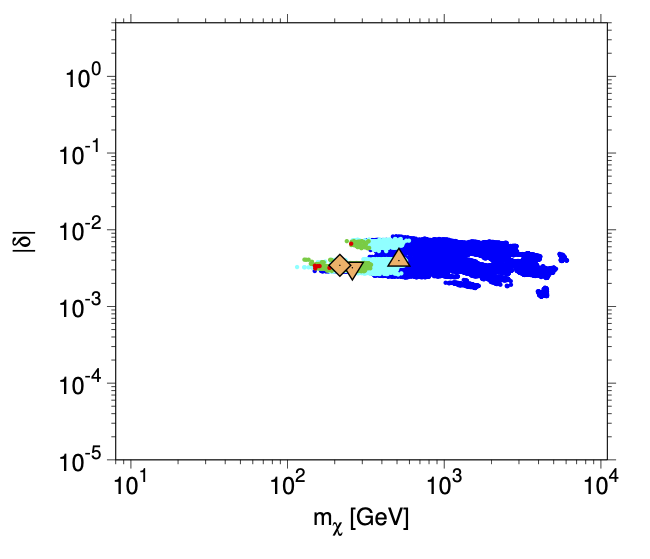}
\includegraphics[width=0.45\textwidth]{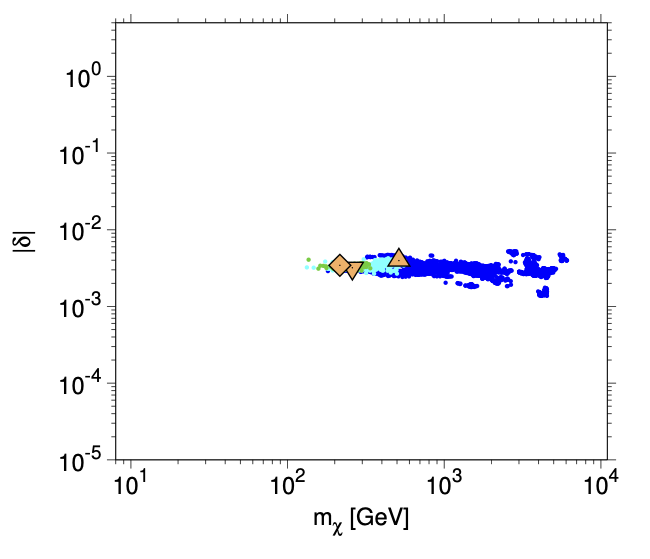}\\
 \caption{\it  The Higgsino components $\gamma$ and $\delta$, respectively, of the LSP for the NUGM points shown in Fig.~\ref{fig:Gaugino_masses} as functions of  $m_\chi$. The color coding corresponds to the values of $\Delta a_\mu$
 shown in Fig.~\ref{fig:Gaugino_masses}.
}
\label{fig:Higgsino_content}
\end{figure}

For completeness, we display in Fig.~\ref{fig:Higgsino_content} the
magnitudes of the Higgsino components in our NUGM sample. We see that these
components are generally small, in the range $10^{-2} - 10^{-3}$, and do
not exhibit much dependence on either $m_\chi$ or $\Delta a_\mu$. Results for the full sample
are shown in the left panels, and for models with $\ohsq$ within the range of cold dark matter density favored by Planck in the right panels.

\begin{figure}[ht!]
\centering
\includegraphics[width=0.45\textwidth]{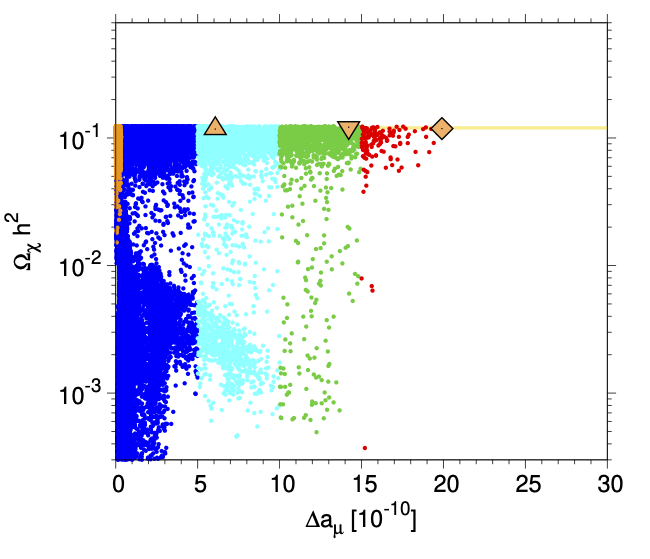}
 \includegraphics[width=0.45\textwidth]{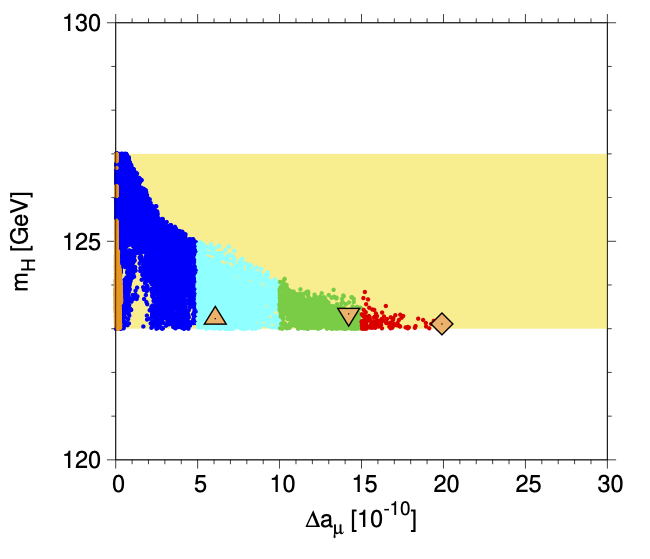}\\
 \vspace{-2mm}
 \includegraphics[width=0.45\textwidth]{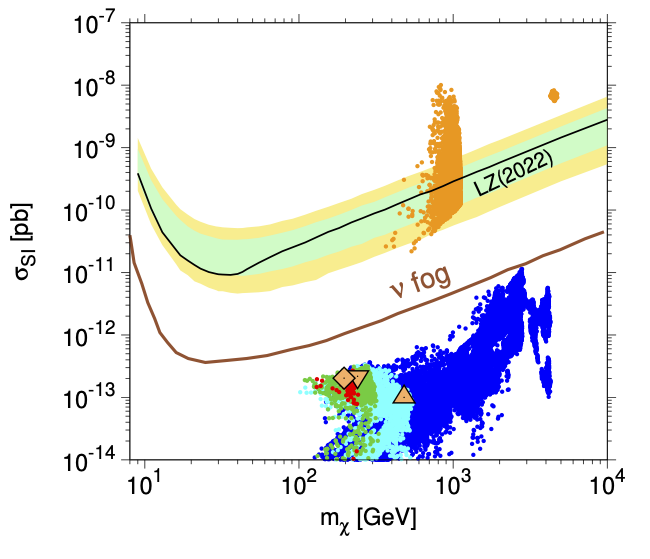} 
  \includegraphics[width=0.45\textwidth]{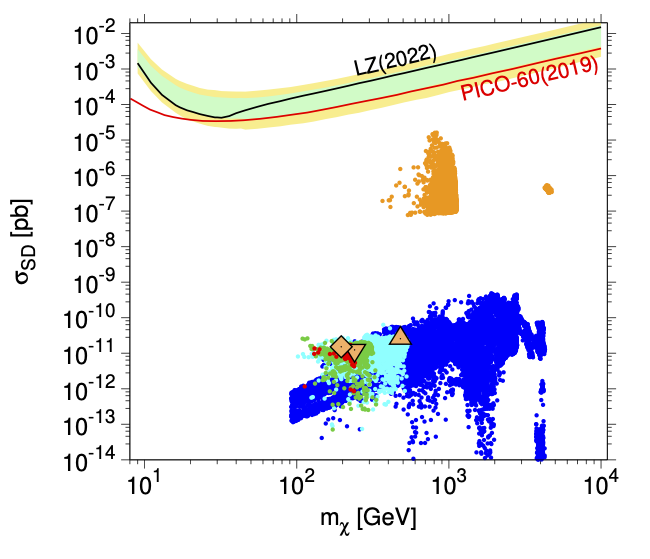}  \\
% \vspace{-2mm}
% \includegraphics[width=0.45\textwidth]{smuon_gm2_scx_c}
%\includegraphics[width=0.45\textwidth]{gm2_mchi_scx_c}\\
\caption{\it 
Scatter plots of projections of the results for the NUGM from MCMC runs 
with non-universal $M_{1,2,3}$ but universal $m_0$ and $A_0$. These parameters and  $\tan\beta$ are allowed to vary over the ranges given in Eq.~(\ref{r1}). We show the relic LSP density, $\Omega_\chi h^2$ (top left); the Higgs mass, $m_H$ (top right); the spin-independent (SI) $\chi-p$ scattering cross section, $\sigma_{\rm SI}$ (lower left) and the spin-dependent (SD) $\chi-p$ scattering cross section, $\sigma_{\rm SD}$ (lower right). Points in these panels were selected by imposing only the upper limit to the LSP density
 $   \Omega_\chi h^2 \le 0.1236$.
 The direct detection cross sections are re-scaled  by a factor $\Omega_{\chi} h^2/0.12 $.
The orange points correspond to a sample with unified gaugino masses (the CMSSM). The other colors correspond to the different $5 \times 10^{-10}$ ranges of $\Delta a_\mu$. {The neutrino fog layer~\cite{fog} for SI scattering is indicated by a brown line.}
 }
\label{fig:gm2_planes}
\end{figure}

\begin{figure}[ht!]
\centering
\includegraphics[width=0.45\textwidth]{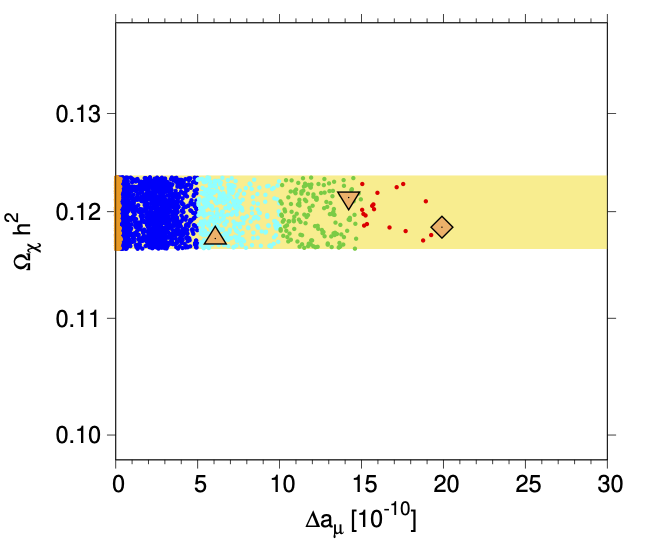}
 \includegraphics[width=0.45\textwidth]{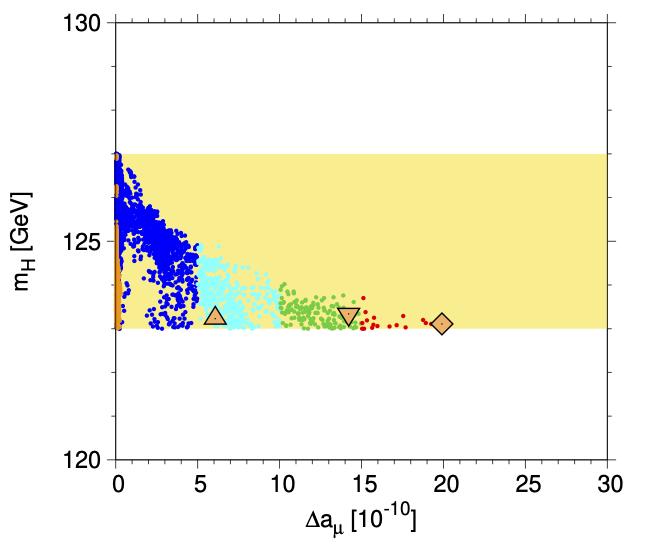}\\
 \vspace{-2mm}
 \includegraphics[width=0.45\textwidth]{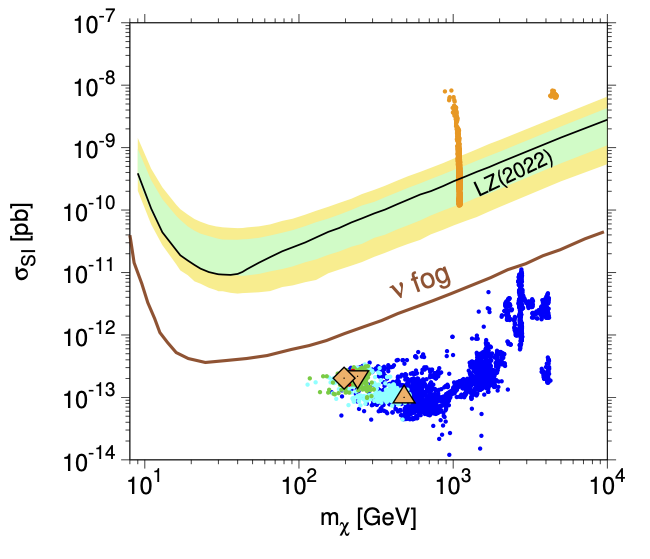} 
  \includegraphics[width=0.45\textwidth]{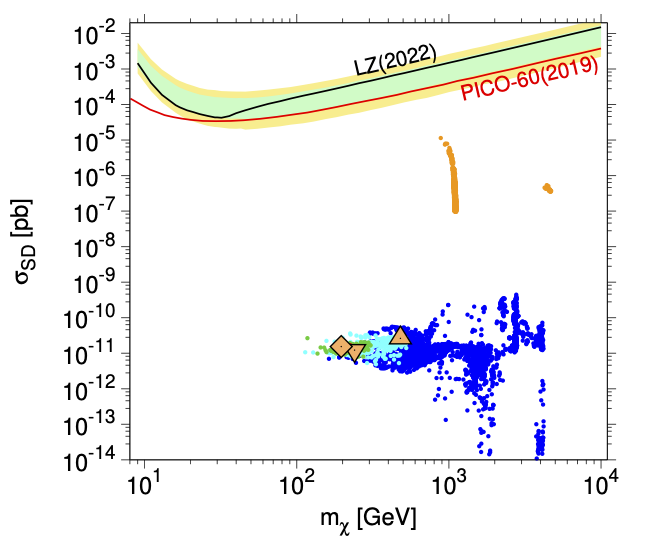}  \\
% \vspace{-2mm}
% \includegraphics[width=0.45\textwidth]{smuon_gm2_sc_c}
%\includegraphics[width=0.45\textwidth]{gm2_mchi_sc_c}\\
\caption{\it 
As in Fig.~\ref{fig:gm2_planes}, but restricting the LSP density
to the preferred cosmological range $  0.1164 \le \Omega_\chi h^2 \le 0.1236$.
}
\label{fig:gm2_planes_CDM}
\end{figure}

The upper left panel of Fig.~\ref{fig:gm2_planes} displays our results for the relic density from the scan of parameters in the
NUGM. These points are color-coded according to their values of $\Delta a_\mu$. We also show in orange
a comparison sample of CMSSM points, for which $M_1 = M_2 = M_3$. We see that these yield only very small values of $\Delta a_\mu$, as expected. We note also in the upper panels of
Fig.~\ref{fig:gm2_planes} that the NUGM points yield only $\Delta a_\mu \lesssim 20 \times 10^{-10}$.
The reason for this is apparent in the upper right panel which shows the value of $m_H$ as a function of $\Delta a_\mu$: larger values of $\Delta a_\mu$ correspond to $m_H$ outside the specified
range. We see in the upper left panel of Fig.~\ref{fig:gm2_planes} that most of the allowed points yield values of $\ohsq$ below the range
specified above. Those points whose relic density are restricted to the range $  0.1164 \le \Omega_\chi h^2 \le 0.1236$ are shown in Fig.~\ref{fig:gm2_planes_CDM}. 
This restriction is immediately apparent in the
upper left panel of Fig.~\ref{fig:gm2_planes_CDM}, but has little impact in the $(\Delta a_\mu, m_H)$ plane shown in the
upper right panel.

The lower panels of
Figs.~\ref{fig:gm2_planes} and \ref{fig:gm2_planes_CDM} show scatter plots of the LSP mass, $m_\chi$, vs the spin-independent (SI) LSP-proton scattering cross section (left)
and the spin-dependent (SD) LSP-proton scattering cross section (right). 
In Fig.~\ref{fig:gm2_planes} the scattering cross sections have been scaled by a factor $\Omega_\chi h^2/0.12$
to compensate for the reduced value of $\Omega_\chi h^2$, but no such scaling is done in Fig.~\ref{fig:gm2_planes_CDM}. We see in all of these panels that the NUGM cross sections are orders of magnitude
below the current experimental upper limits, below the estimated level of the neutrino `fog layer'~\cite{fog} except for some points with $\Delta a_\mu < 5 \times 10^{-10}$,
whereas the CMSSM can yield values of the SI LSP-proton scattering cross section above the
corresponding experimental limit, and values of the SD LSP-proton scattering cross section that are only slightly below the experimental limit. Note that we do not show the neutrino fog line in the spin-dependent case as it is highly dependent on the detector material used \cite{fog}.

\begin{figure}[ht!]
\centering
\includegraphics[width=0.45\textwidth]{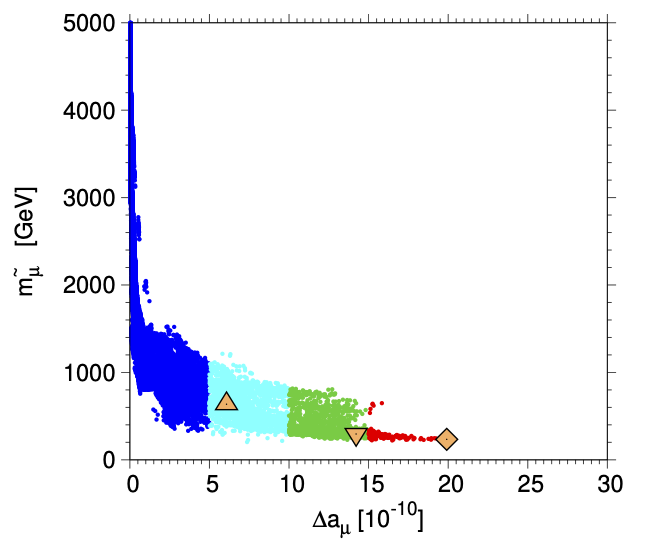}
\includegraphics[width=0.45\textwidth]{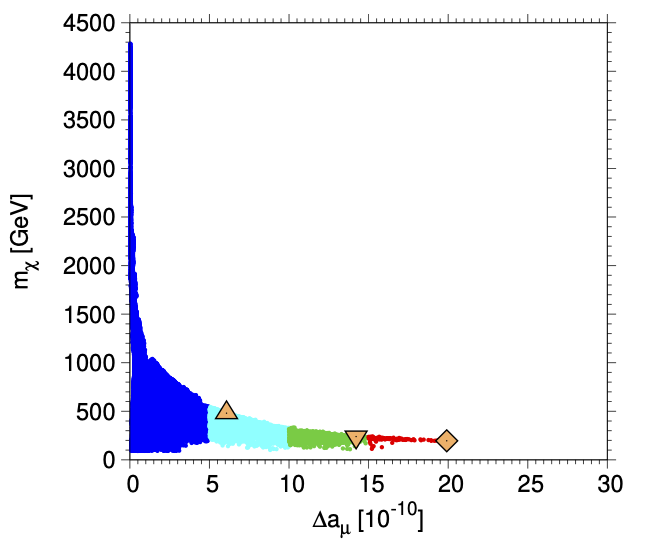}\\
\includegraphics[width=0.45\textwidth]{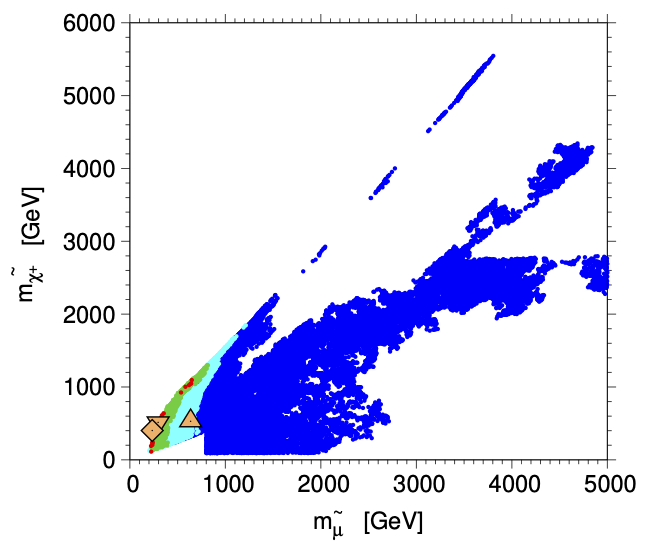}
 \includegraphics[width=0.45\textwidth]{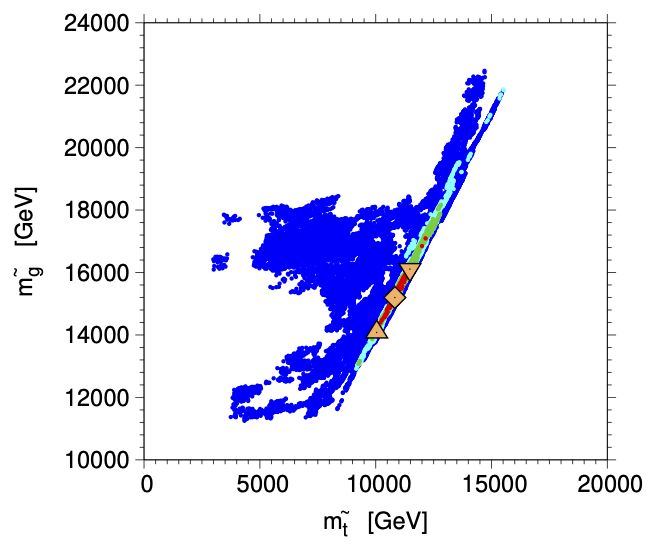} \\
\caption{\it As in Fig.~\ref{fig:gm2_planes}, displaying the allowed values of
sparticle masses in the NUGM as functions of $\Delta a_\mu$.
Only the upper limit on the cosmological range $ \Omega_\chi h^2 \le 0.1236$ is imposed.}
\label{fig:NUGM_masses_CDM}
\end{figure}

Figure~\ref{fig:NUGM_masses_CDM} displays predictions for some sparticle masses in the NUGM,
using the same color coding as in Fig.~\ref{fig:Gaugino_masses}. We see in the upper left panel how the range of smuon masses is correlated
with the value of $\Delta a_\mu$, and in the upper right panel we see similar behavior for the LSP mass, $m_\chi$. As expected the
lower smuon and LSP masses are correlated with larger values of $\Delta a_\mu$. The correlation between the masses of the smuon and the lighter (Wino-like) chargino, $\chi^\pm$,
is shown in the lower left panel of Fig.~\ref{fig:NUGM_masses_CDM}, and the correlation between the stop and gluino masses is shown
in the lower right panel.~\footnote{Points in this figure are only subject to the upper limit on the relic density. 
Restricting to the narrow range of relic densities would only thin the density of points shown, but has little impact on the correlations of $m_{\tilde \mu}$ and $m_\chi$ with $\Delta a_\mu$.}
We note that large values of $\Delta a_\mu$ correspond to large values of these masses, well beyond the reach
of LHC experiments. In the NUGM, 
while the smuon mass is restricted
to $\lesssim 200$~GeV if $\Delta a_\mu > 15 \times 10^{-10}$, we find that $m_{\tilde t} > 11$~TeV and $m_{\tilde g} \sim 15$~TeV.
These results illustrate how the NUGM can reconcile a relatively large value of $\Delta a_\mu$ with the favored range
of $\ohsq$ and the theoretical calculation of $m_H$. This reconciliation requires, in addition, 
 that $\mu \sim 8$~TeV.

\subsection{Non-Universal Higgs and Third-Generation Scalar Mass Model (NUHM3)}
\label{sec:NUHM3}

Our analysis of the model with non-universal Higgs and third-generation scalars
(NUHM3) proceeds along similar lines to the case of the NUGM analysis above.
In the NUHM3 case the scan ranges for the seven model parameters are~\footnote{As noted earlier, the choice of $\mu$ and $M_A$ is equivalent to using the soft supersymmetry-breaking masses $m_{H_1}$ and $m_{H_2}$. Using the former allows a more efficient search of parameters sets making a sizeable contribution to $\Delta a_\mu$. } 
\begin{align}
M_{1/2} & = 0 - 6 \TeV \, , \nonumber \\
m_{012} & =  0 - 6 \TeV \, , \nonumber \\
m_{03} & =  0 - 20  \TeV \, , \nonumber \\
\mu & = 0 - 5 \TeV \, ,  \nonumber \\
M_A & = 0 - 20 \TeV  \, , \nonumber \\
|A_0/m_0| & = 0 - 40 \, , \nonumber \\
\tan \beta & = 1 - 50 \, .
\end{align}
As in the NUGM case, the supersymmetry-breaking mass parameters are input at the GUT
scale and run down to the electroweak scale. Inputs that do not yield a neutralino LSP,
lead to a tachyonic Higgs pseudoscalar, or do not satisfy the electroweak symmetry-breaking minimization conditions are discarded.

In our scan of the seven-dimensional parameter space of the NUHM3, the MCMC chains provide us with 600,000 points initially. Some 100,000 of them pass the kinematical constraints and have a neutralino LSP. Of these, about 60,000 points satisfy the  conditions 
$m_H=125 \pm 2 \GeV$ and  $\Omega_\chi h^2 \leq 0.1236$, and about 5,000 points satisfy the stricter condition 
$0.1164 \leq \Omega_\chi h^2 \leq 0.1236$. 

\begin{figure}[ht!]
\centering
\includegraphics[width=0.45\textwidth]{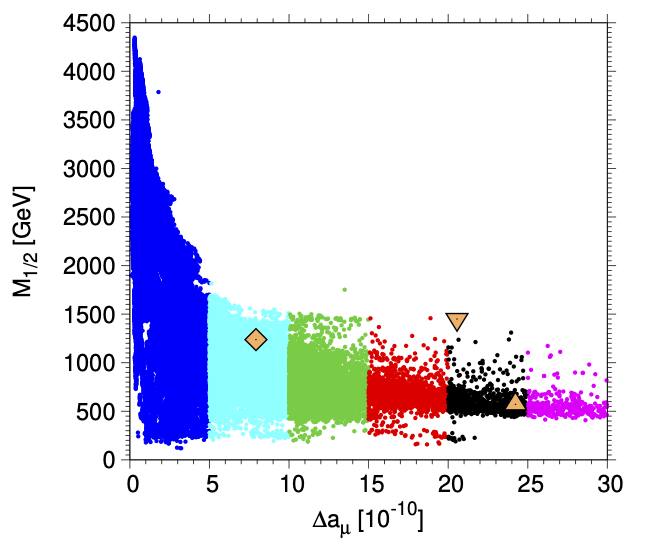}
\includegraphics[width=0.45\textwidth]{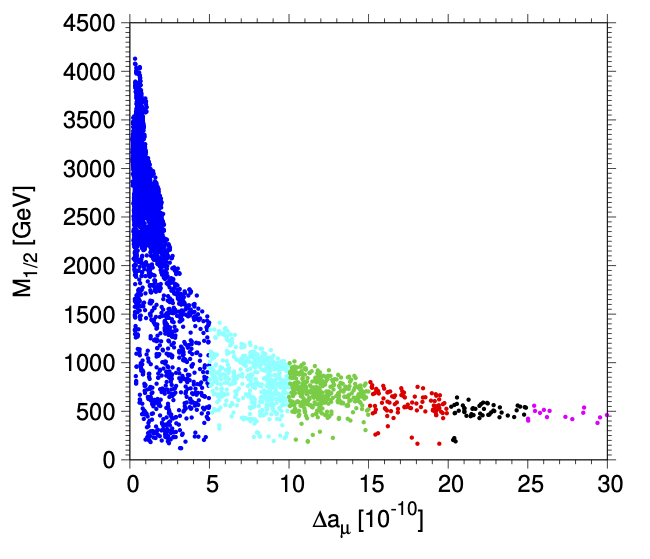}\\
 \vspace{-2mm}
\includegraphics[width=0.45\textwidth]{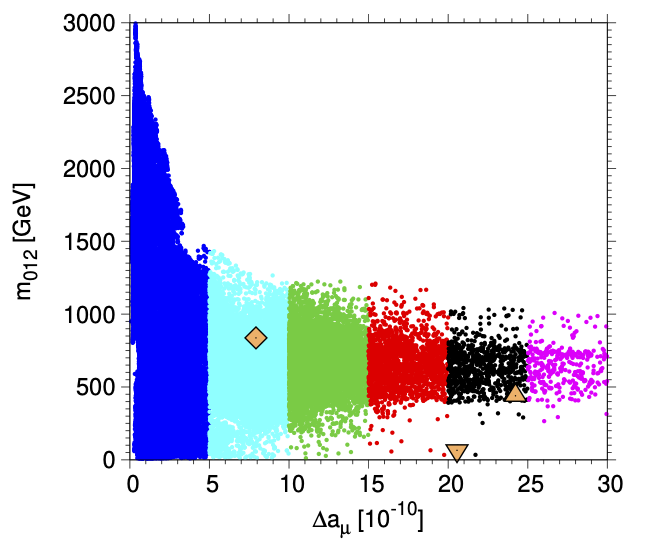}
\includegraphics[width=0.45\textwidth]{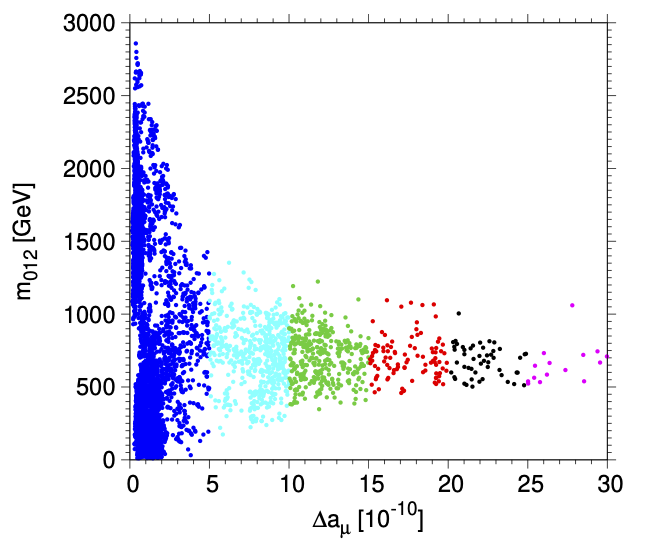}\\
 \vspace{-2mm}
\includegraphics[width=0.45\textwidth]{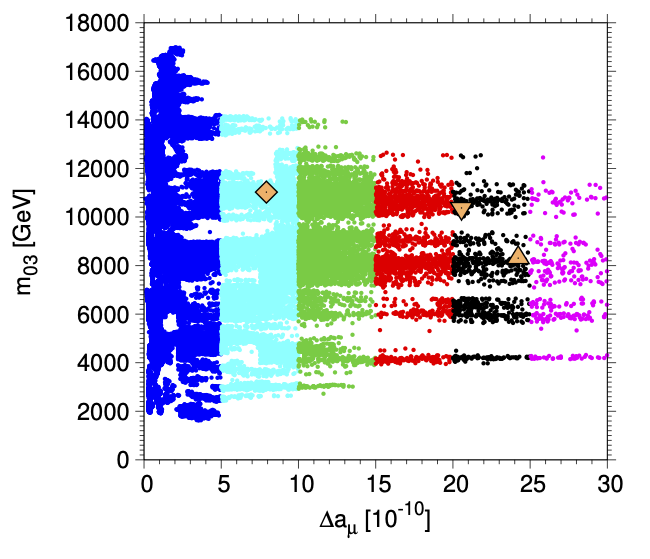}
\includegraphics[width=0.45\textwidth]{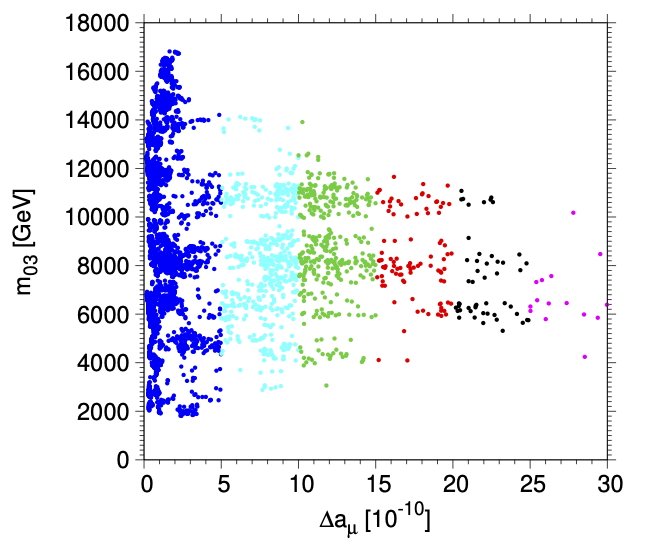}\\
\caption{\it  The allowed values of
$M_{1/2}, m_{012}$ and $m_{03}$ in the NUHM3 model as functions of $\Delta a_\mu$, for points with with $ \Omega_\chi h^2 \le 0.1236$ on the left and points with $0.1164 \le \Omega_\chi h^2 \le 0.1236$ on the right. {The triangles in the left planes indicate the locations of the NUHM3 benchmark points discussed in Section~\ref{sec:benchmarks} below.}
}
\label{fig:NUHM3_M12_m0}
\end{figure}

\begin{figure}[ht!]
\centering
\includegraphics[width=0.45\textwidth]{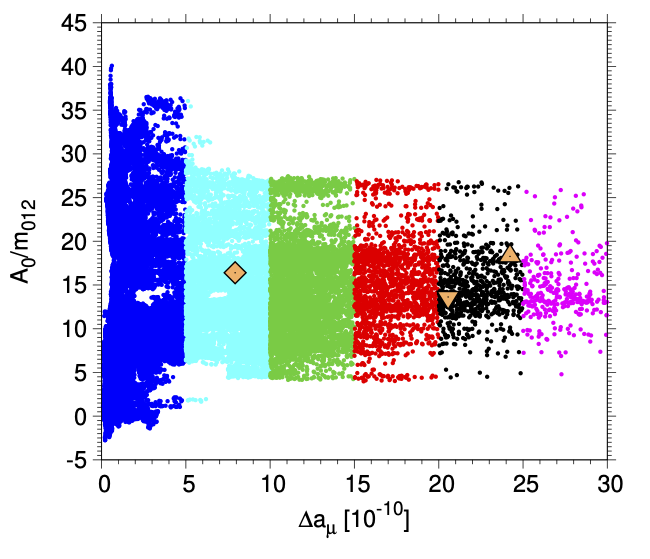}
\includegraphics[width=0.45\textwidth]{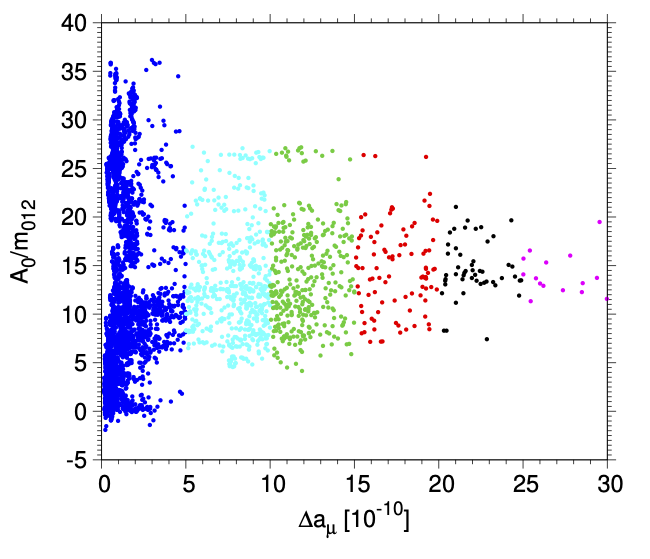} \\
 \includegraphics[width=0.45\textwidth]{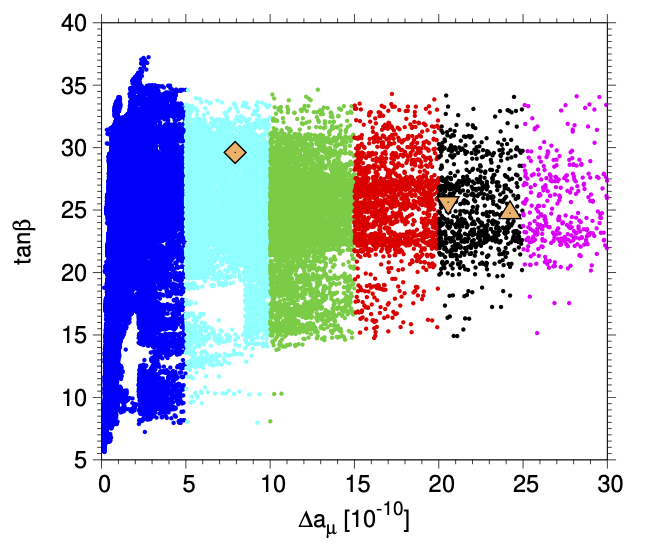}
 \includegraphics[width=0.45\textwidth]{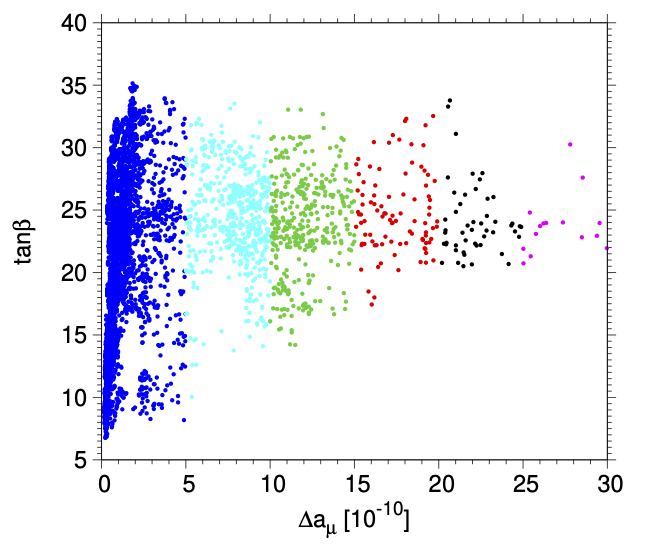} \\
\caption{\it As in Fig.~\ref{fig:NUHM3_M12_m0}, displaying the allowed values of
$A_0/m_{012}$ and $\tan \beta$ in the NUHM3 model as functions of $\Delta a_\mu$, for points with with $ \Omega_\chi h^2 \le 0.1236$ on the left and points with $0.1164 \le \Omega_\chi h^2 \le 0.1236$ on the right.}
\label{fig:NUHM3_A0tb}
\end{figure}

\begin{figure}[ht!]
\centering
 \includegraphics[width=0.45\textwidth]{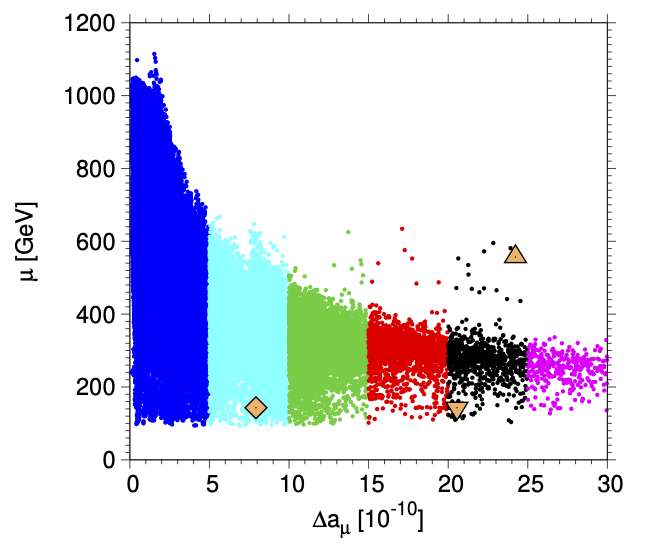}
 \includegraphics[width=0.45\textwidth]{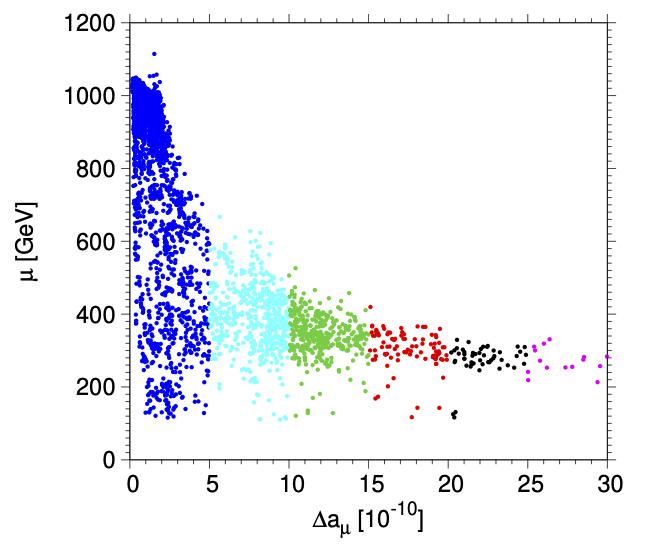} \\
\includegraphics[width=0.45\textwidth]{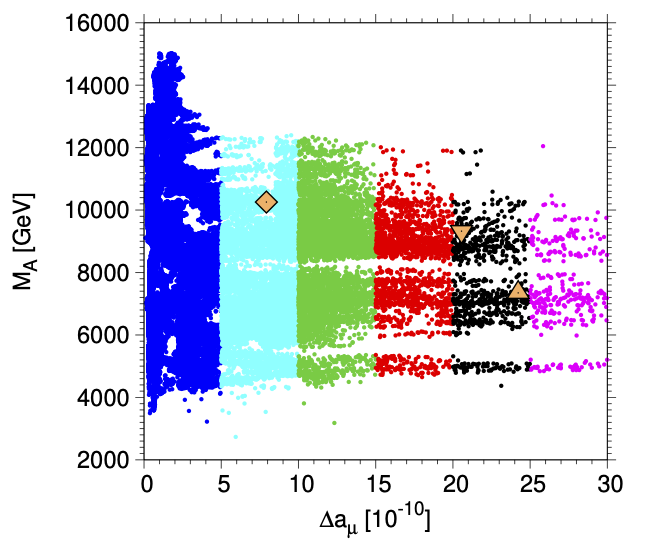}
\includegraphics[width=0.45\textwidth]{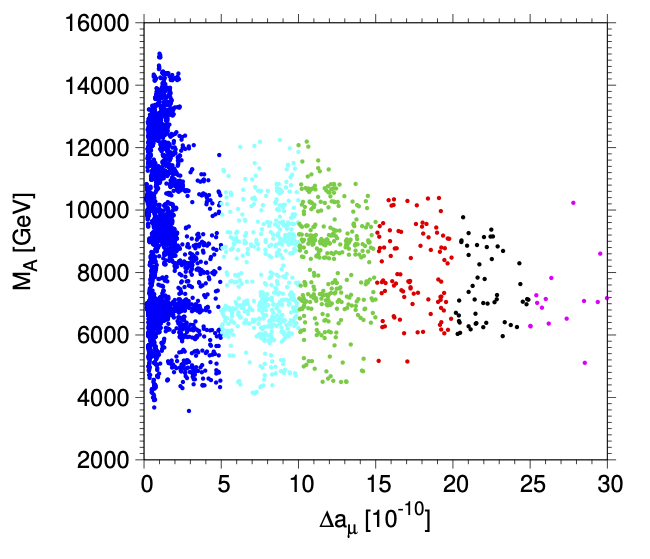} \\
\caption{\it As in Fig.~\ref{fig:NUHM3_M12_m0}, displaying the allowed values of
$\mu$ and $M_A$ in the NUHM3 model as functions of $\Delta a_\mu$, for points with with $ \Omega_\chi h^2 \le 0.1236$ on the left and points with $0.1164 \le \Omega_\chi h^2 \le 0.1236$ on the right.}
\label{fig:NUHM3_muMA}
\end{figure}

We show in Figs.~\ref{fig:NUHM3_M12_m0}, \ref{fig:NUHM3_A0tb}, and \ref{fig:NUHM3_muMA}
the values of the NUHM3 input parameters for points that survive 
all of the constraints discussed above. As for the NUGM, 
these are color-coded in intervals of 
$\Delta a_\mu = 5 \times 10^{-10}$.
We see that values of $\Delta a_\mu$ as large as $30 \times 10^{-10}$
are possible, but for relatively few points when the LSP density is
restricted to the Planck determination of $\ohsq$ (right panels). 
The parameter sets yielding large values of $\Delta a_\mu$ tend to prefer relatively 
small values of the common gaugino mass $M_{1/2}$.
As we see from the top right panel of Fig.~\ref{fig:NUHM3_M12_m0},
$M_{1/2} \simeq 500$~GeV when the LSP makes up all of the dark matter. Similarly, 
when $\Delta a_\mu$ is large the common 
first- and second-generation scalar mass $m_{012}$ also takes low values with $m_{012} \simeq 500-700$~GeV preferred. In contrast, the spread
third-generation scalar mass $m_{03}$ is generally much larger, so as to accommodate a large stop mass as needed to provide an acceptable value of $m_H$.
While the data appear to select somewhat distinct values of $m_{03}$
at large $\Delta a_\mu$ in the bottom left panel of Fig.~\ref{fig:NUHM3_M12_m0}, this is due to limitations of the MC sampling, and does not affect the overall envelope of allowed points in the $(\Delta a_\mu, m_{03})$ plane.
{The triangles in the left  planes indicate the locations of the NUHM3 benchmark points discussed in Section~\ref{sec:benchmarks} below: none of the NUHM3 benchmarks saturate the Planck cold dark
matter density.}

Fig.~\ref{fig:NUHM3_A0tb} indicates that the NUHM3 models are relatively insensitive to $A_0$ and $\tan \beta$. There is a wide range of values of $A_0/m_{012}$ that yield large values of $\Delta a_\mu$, though the number of points found is greatly diminished when we restrict the density to the cosmological range as we see when comparing the left and right panels.
We note that although the values of $A_0/m_{012}$ are large, namely $\mathcal{O}(20)$), $A_0/m_{03}$ is only $\mathcal{O}(1)$. Compared to the NUGM discussed in the previous section, values of $\tan \beta$ found here are relatively high, of order 25. 

In contrast to $A_0/m_{012}$, the value of $\mu$ shown in Fig.~\ref{fig:NUHM3_muMA} does correlate with $\Delta a_\mu$, with lower values of $\mu$ preferred for large $\Delta a_\mu$.  When the LSP density is restricted to the cosmological range, $\mu \lesssim 300$~GeV is preferred as seen in the upper right panel of Fig.~\ref{fig:NUHM3_muMA}. On the hand, a large range of values of $M_A$ between 5 and 10 TeV are found at large when $\Delta a_\mu$ is large and the LSP density is not restricted. With the relic density restricted, there are again fewer points at large $\Delta a_\mu$.

\clearpage

\begin{figure}[ht!]
\centering
\includegraphics[width=0.45\textwidth]{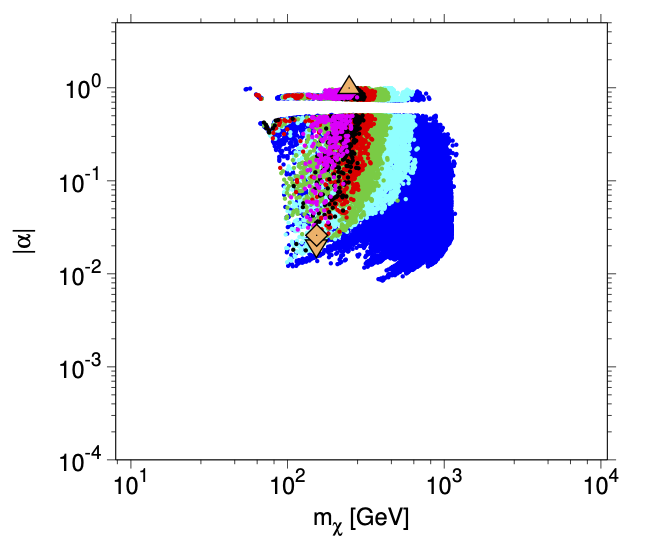}
\includegraphics[width=0.45\textwidth]{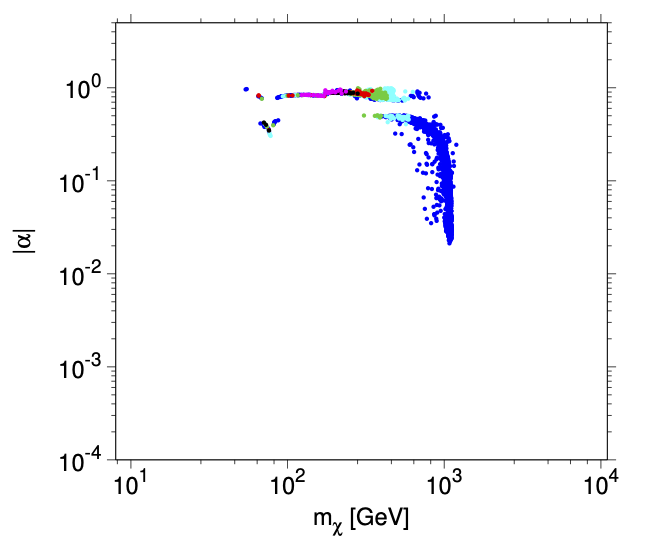}\\
  \vspace{-2mm}
\includegraphics[width=0.45\textwidth]{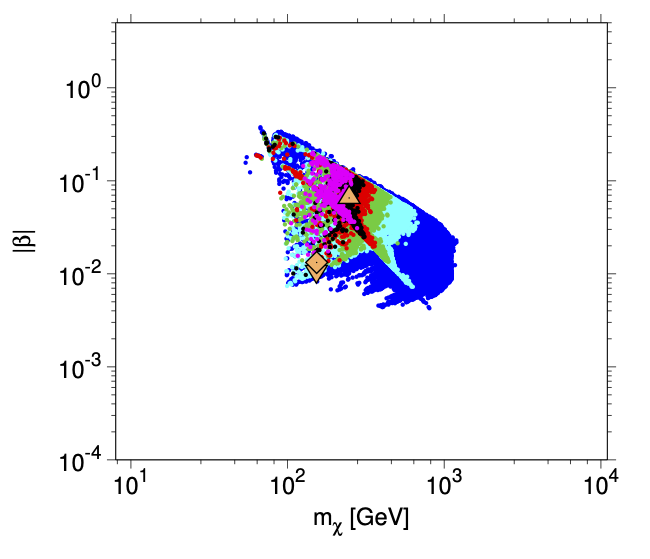}
\includegraphics[width=0.45\textwidth]{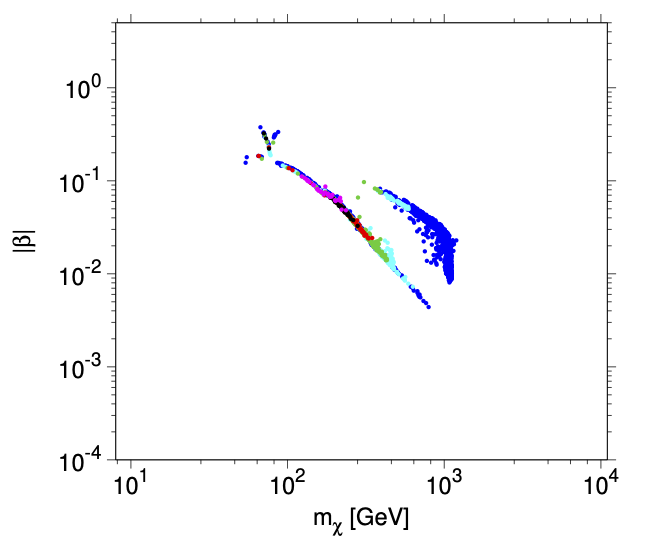}\\
 \caption{\it  The Bino (upper panels) and Wino  (lower panels) components $\alpha$ and $\beta$, respectively, of the LSP for the NUHM3 points shown in Fig.~\ref{fig:NUHM3_M12_m0} as functions of  $m_\chi$. The color coding corresponds to the values of $\Delta a_\mu$
 shown in Fig.~\ref{fig:NUHM3_M12_m0}.}
\label{fig:NUHM3_Gaugino_content}
\end{figure}

\begin{figure}[ht!]
\centering
\includegraphics[width=0.45\textwidth]{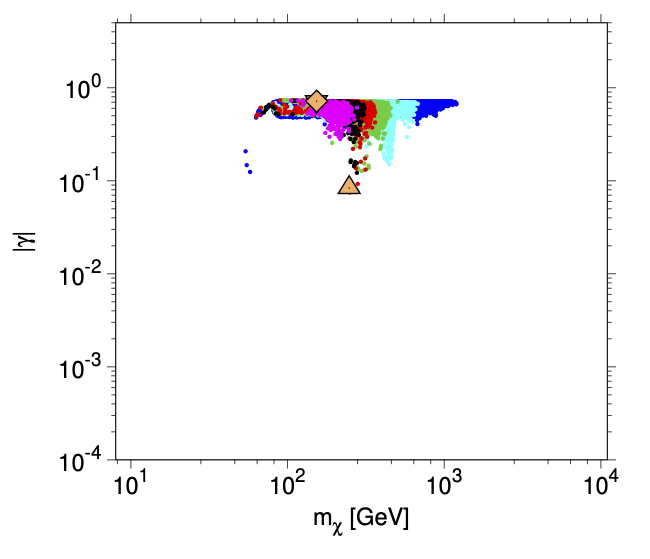} 
\includegraphics[width=0.45\textwidth]{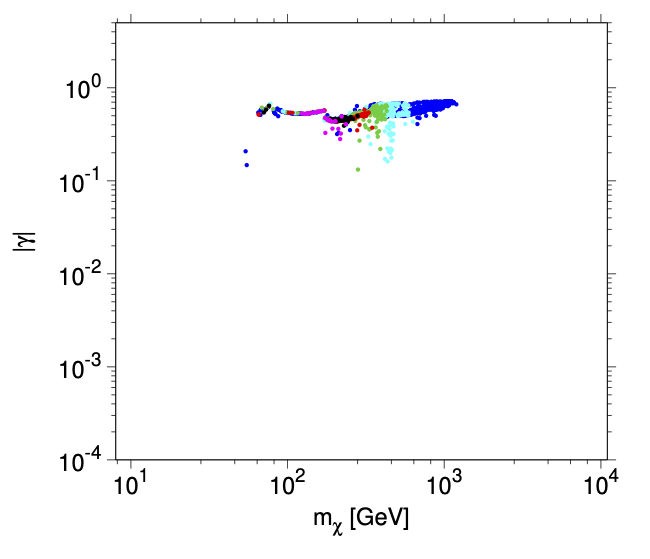}\\
  \vspace{-2mm}
\includegraphics[width=0.45\textwidth]{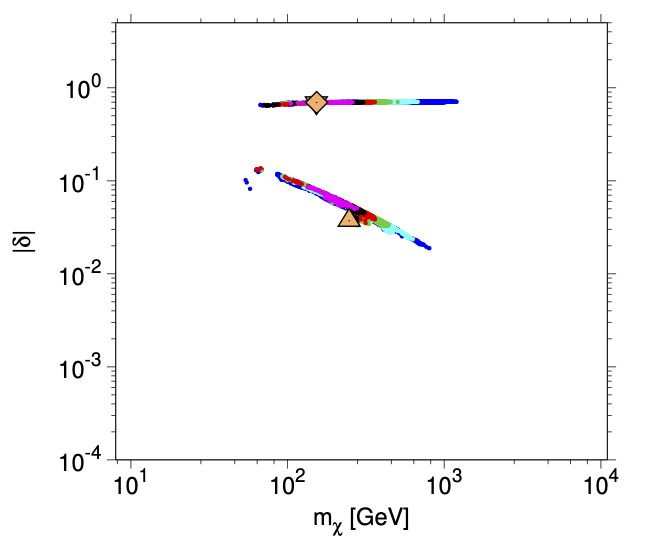}
\includegraphics[width=0.45\textwidth]{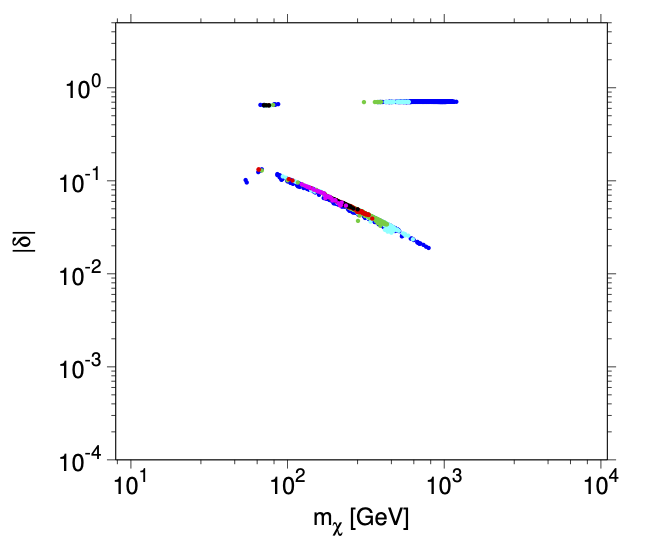}\\
 \caption{\it  The Higgsino components $\gamma$ and $\delta$, respectively, of the LSP for the NUHM3 points shown in Fig.~\ref{fig:NUHM3_M12_m0} as functions of  $m_\chi$. The color coding corresponds to the values of $\Delta a_\mu$
 shown in Fig.~\ref{fig:NUHM3_M12_m0}. Though difficult to see, benchmark points B and C are sitting on top of each other in this figure.
}
\label{fig:NUHM3_Higgsino_content}
\end{figure}

We display in Fig.~\ref{fig:NUHM3_Gaugino_content}
the composition of the LSP in these NUHM3 points. 
The left panels display points with $\Omega_\chi h^2 \le 0.1236$ and the right panel shows points
with $  0.1164 \le \Omega_\chi h^2 \le 0.1236$. In both cases $m_H = 125\pm 2 \GeV$.
Unlike the case of the
the NUGM, the Bino component $\alpha$ may or may not dominate.
Indeed, we find two distinct populations, those with $|\alpha| \sim 1$ and those with much lower values, for which one of the Higgsino components dominates.  The two Higgsino components, $\gamma$ and $\delta$ are displayed in Fig.~\ref{fig:NUHM3_Higgsino_content}, again for points with $\Omega_\chi h^2 \le 0.1236$ in the left panels and for points
with $  0.1164 \le \Omega_\chi h^2 \le 0.1236$ in the right panels, and $m_H = 125\pm 2 \GeV$.
The Wino component $\beta$ is mostly in the range $[0.1, 0.01]$
though values as large as $\sim 0.3$ are attained for a few points.

\begin{figure}[ht!]
\centering
\includegraphics[width=0.45\textwidth]{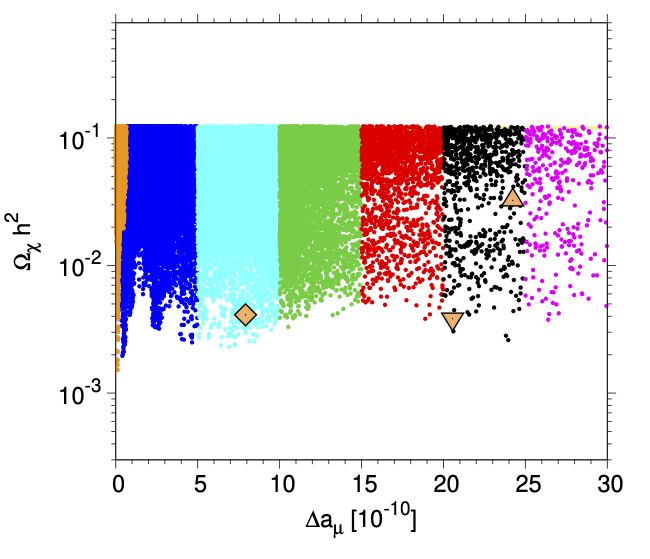}
\includegraphics[width=0.45\textwidth]{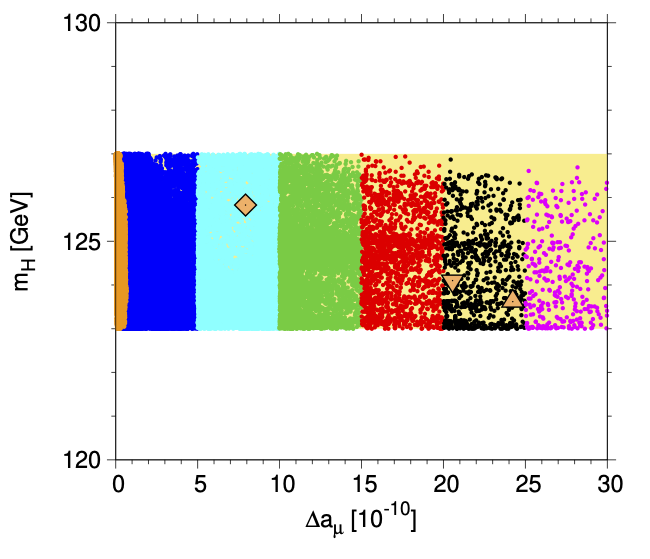}\\
 \vspace{-2mm}
\includegraphics[width=0.45\textwidth]{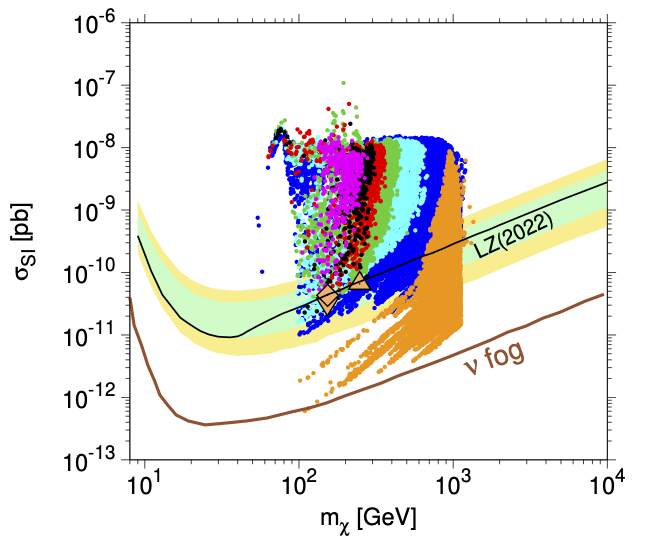} 
\includegraphics[width=0.45\textwidth]{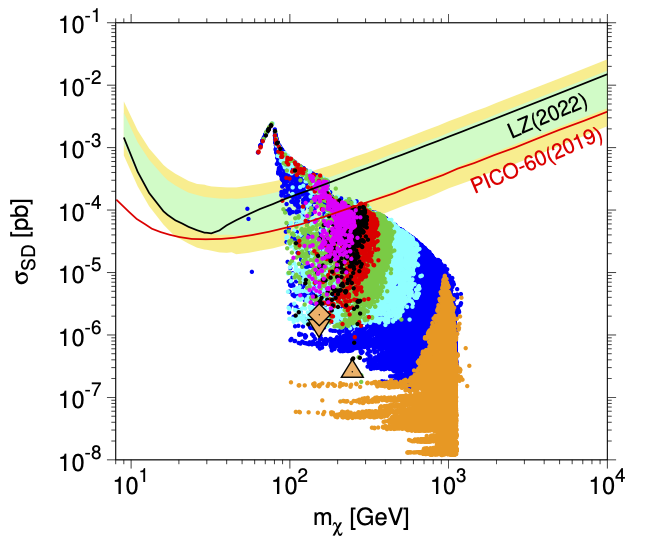}  \\
% \vspace{-2mm}
% \includegraphics[width=0.45\textwidth]{smuon_gm2_scx_c}
% \includegraphics[width=0.45\textwidth]{gm2_mchi_scx_c}\\
\caption{\it  As in Fig.~\ref{fig:gm2_planes}, but for the NUHM3 model, allowing  $M_{1/2}$, $m_{012}$,  $m_{03}$, $\mu$, $M_A$,  $A_0$ and  $\tan\beta$ to vary, showing all points with $\Omega_\chi h^2 \le 0.1236$ and rescaling the direct detection cross sections by a factor $\Omega_{\chi} h^2/0.12 $.
The orange  points correspond to a sample with  $m_{012}$=$m_{03}$, i.e., the NUHM2 model.}
\label{fig:NUHM3_gm2_planes}
\end{figure}

Figure~\ref{fig:NUHM3_gm2_planes}  displays some two-dimensional
projections of the NUHM3 points allowed when the relic LSP density is constrained only by the upper limit
$\Omega_\chi h^2 \le 0.1236$. 
Here, in addition to the color coding for different 
$5 \times 10^{-10}$ ranges of $\Delta a_\mu$, we have colored orange
points in a sample with $m_{012}$=$m_{03}$ (the NUHM2). As already
commented, there are NUHM3 points compatible with $m_H =125 \pm 2$~GeV
that have $\Delta a_\mu$ as large as $30 \times 10^{-10}$, whereas the
NUHM2 points all have very small $\Delta a_\mu$, as was the case for the
CMSSM sample mentioned in connection with our NUGM analysis above.
The scattering cross sections are shown in the lower panels of Fig.~\ref{fig:NUHM3_gm2_planes}.
In this case, as in the NUGM, we plot the scattering cross section scaled by the density $\Omega_\chi h^2/0.12$, so as to correspond better to the scattering rate in a detector. 
There is wide range of the SI
and SD cross sections that are allowed by experiment for points with a low density. Measurements of the SI cross
section exclude most of the NUHM3 sample, but some points with
$\Delta a_\mu \lesssim 15 \times 10^{-10}$ are allowed, along with
many of the NUHM2 points. The present experimental limit of the SD
cross section is a weaker constraint on the NUHM3 sample, and all the
NUHM2 sample predicts values well below the experimental limit.

\begin{figure}[ht!]
\centering
\includegraphics[width=0.45\textwidth]{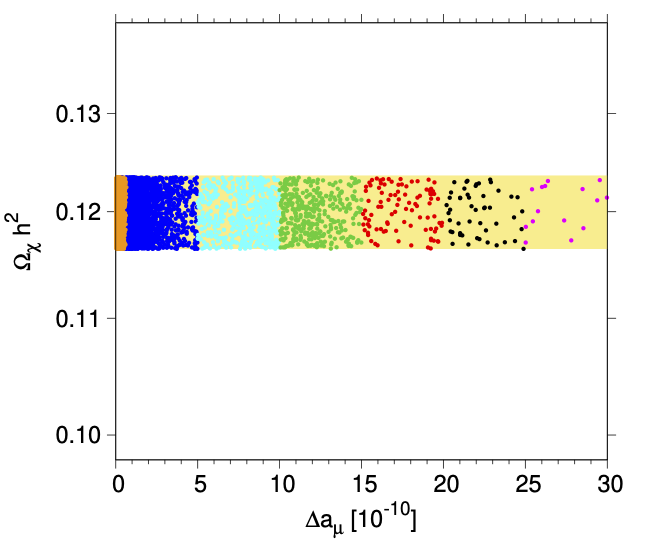}
\includegraphics[width=0.45\textwidth]{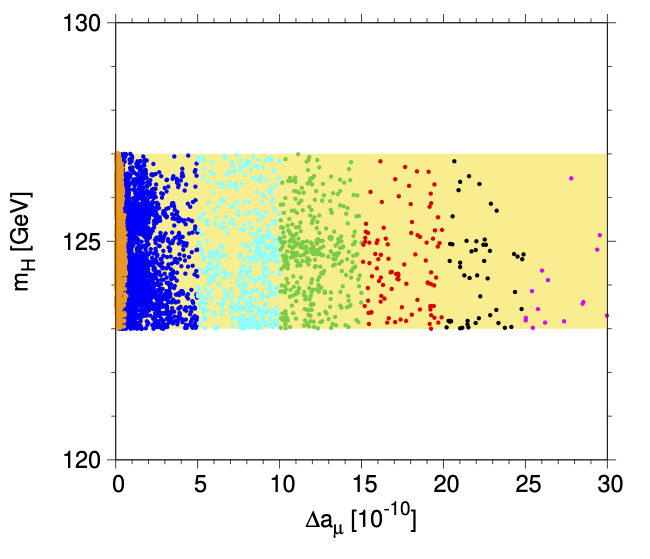}\\
 \vspace{-2mm}
\includegraphics[width=0.45\textwidth]{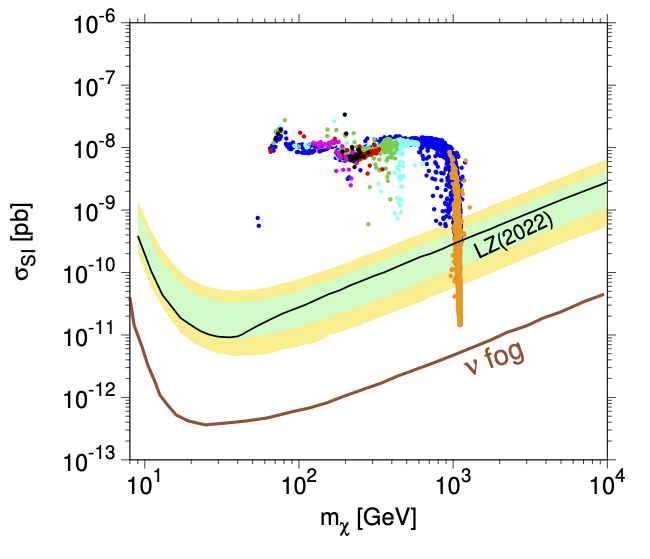} 
\includegraphics[width=0.45\textwidth]{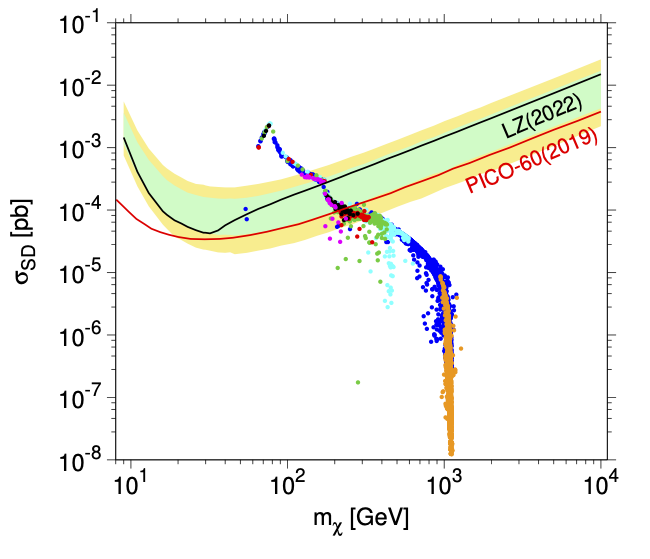}  \\
% \vspace{-2mm}
% \includegraphics[width=0.45\textwidth]{smuon_gm2_scx_c}
% \includegraphics[width=0.45\textwidth]{gm2_mchi_scx_c}\\
\caption{\it As in Fig.~\ref{fig:NUHM3_gm2_planes}, but restricting the LSP density to the Planck range $0.1164 \le \Omega_\chi h^2 \le 0.1236$.}
\label{fig:NUHM3_gm2_planes_0hsq}
\end{figure}

Figure~\ref{fig:NUHM3_gm2_planes_0hsq} shows an analogous set of NUHM3
parameter planes where the LSP density is
restricted to the Planck range $  0.1164 \le \Omega_\chi h^2 \le 0.1236$.
We see that
there are still many such points with $m_H$ within $\pm 2$~GeV of the measured Higgs mass. 
The lower left panel of Fig.~\ref{fig:NUHM3_gm2_planes_0hsq} shows, however, that the NUHM3 sample points generally have values of the SI 
cross section that are excluded, with the remainder lying above the neutrino `fog layer'~\cite{fog}, and that some of the NUHM3 sample
also predicts values of the SD cross section that are excluded.
This is why none of the NUHM3 benchmarks saturate the Planck cold dark matter density.
However, some (all) of the NUHM2 sample predicts values of the SI
(SD) cross section that are compatible with experiment.

%\clearpage

\begin{figure}[ht!]
\centering
 \includegraphics[width=0.45\textwidth]{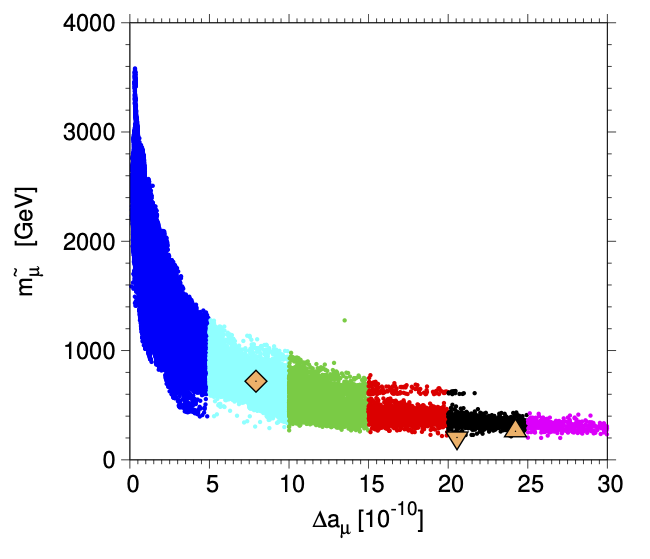}
\includegraphics[width=0.45\textwidth]{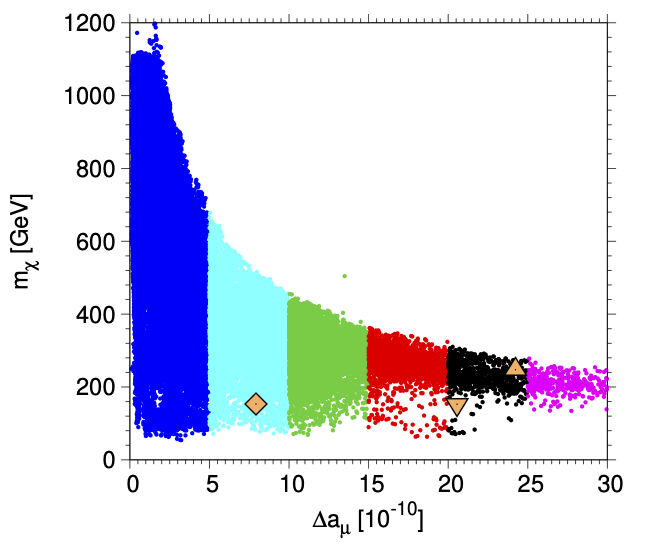}\\
\includegraphics[width=0.45\textwidth]{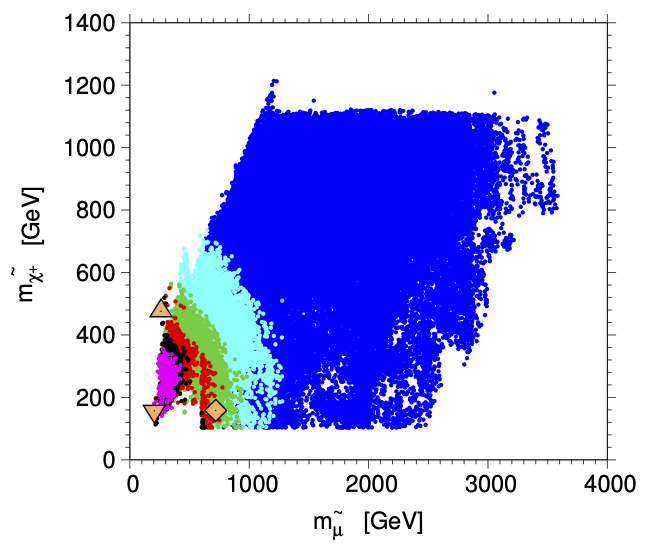}
 \includegraphics[width=0.45\textwidth]{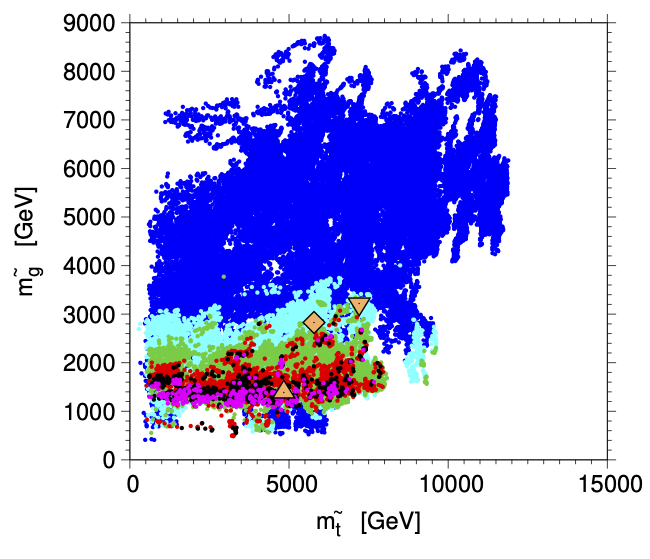} \\
\caption{\it  As in Fig.~\ref{fig:NUGM_masses_CDM}, but displaying the allowed ranges of
sparticle masses in the NUHM3  model as functions of $\Delta a_\mu$.}
\label{fig:NUHM_masses_planes}
\end{figure}

Figure~\ref{fig:NUHM_masses_planes} displays the allowed ranges of 
$m_{\tilde \mu}$ and $m_\chi$ as functions of $\Delta a_\mu$ in the 
upper panels, and the  $(m_{\tilde \mu}, m_{\chi^\pm})$ and $(m_{\tilde t}, m_g)$
correlations in the lower panels, all for the sample with 
$\Omega_\chi h^2 \le 0.1236$. We see that points with
$\Delta a_\mu > 20 \times 10^{-10}$ have $m_{\tilde \mu}$ and $m_\chi$
$\lesssim 300$~GeV, whereas larger masses are allowed for models with
smaller $\Delta a_\mu$. The $(m_{\tilde \mu}, m_\chi)$ correlation is
tighter for points with $\Delta a_\mu > 5 \times 10^{-10}$, but the
larger values of $m_{\tilde \mu}$ and $m_\chi$ allowed for smaller
$\Delta a_\mu$ are largely uncorrelated. We also note that $m_{\tilde g}$
can be large when $\Delta a_\mu > 5 \times 10^{-10}$, but is restricted to
$m_{\tilde g} \lesssim 3$~TeV for larger $\Delta a_\mu$. Qualitatively similar features
remain when we impose $0.1164 \le \Omega_\chi h^2 \le 0.1236$, albeit with a smaller sample.

\clearpage

\section{Benchmark Scenarios}
\label{sec:benchmarks}

Based on the analysis in Section~\ref{sec:NUGM}, we have selected three benchmark NUGM models, labeled A, B, and C. The sets of input parameters for these benchmarks is given in Table \ref{table:bp1}. They are distinguished primarily by the value of $\Delta a_\mu$. The calculated relic density, value of $\Delta a_\mu$, and the spin-independent and -dependent scattering cross sections as well as some representative masses are provided in Table \ref{table:bp1_d}.
Each of these points was chosen to yield a relic density close to the Planck value of the cold dark matter density.

\begin{table}[ht!]
\begin{center}
$\begin{array}{ |c|c|c|c|c|c|c|   }
\hline
{} & M_1 & M_2  &  M_3  &  m_0  &   A_0/m_0  &  \tan\beta     \\
\hline
\hline
%% A  \quad ( \bigtriangleup )  &   1217.74  &   798.71    &   7245.39   &   675.46   &    0.37     &  16.90       \\
A  \quad ( \bigtriangleup )  &   1218  &   799   &   7245   &   676   &    0.4    &  17       \\
\hline
%B \quad ( \bigtriangledown )  &  730.39      &     827.79       &    8364.99      &   541.33    &      -0.59        &    6.78  
B \quad ( \bigtriangledown )  &  730     &     828     &    8365     &   541    &      -0.6      &    7     \\
\hline
%C  \quad ( \Diamond )  &       625.86      &    694.98   &     7852.27       &      549.84        &    0.04      &   7.33     \\
C  \quad ( \Diamond )  &       626      &    695   &     7852       &      550       &    0.0      &   7    \\
\hline
\end{array}$
\caption{\it Benchmark points for the NUGM.}
\label{table:bp1}
\end{center}
\end{table}

\begin{table}[ht!]
\begin{center}
$
\begin{array}{ |c|c|c|c|    }
\hline
{}  & A \quad ( \bigtriangleup )  &  B  \quad ( \bigtriangledown )  &  C  \quad ( \Diamond )       \\
\hline
\hline
\Omega_{\chi} h^2  & 0.118  &    0.123    &   0.118        \\
\hline
\Delta a_{\mu} \, (10^{-10})  &    6.1    &    14.2      &   19.9           \\
\hline
\sigma_{SI} \, (10^{-13}\,  pb)  &  1.02       &   2.16         &    2.06           \\
\hline
 \sigma_{SD} \, (10^{-11}\, pb)  &  2.62       &    1.23         &    1.54            \\
\hline
\hline
 m_\chi  &  480       &    240       &    196             \\
\hline
m_{\chi^+_1}  &   529      &    513        &   402            \\
\hline
m_{\tilde{\mu}_{L,R}}  &  637      &    296       &   234            \\
\hline
m_{\tilde{\tau}_1}  &   513      &    258      &    214            \\
\hline
m_{\tilde{g}}  &  14080      &    16110       &    15200            \\
\hline
m_{\tilde{t_1}}  &   10040      &   11480        &    10830             \\
\hline
\end{array}
$
\caption{\it Observables and masses for the NUGM benchmark points in Table~\ref{table:bp1}. The LSP is Bino-like for these points.}
\label{table:bp1_d}
\end{center}
\end{table}

As seen in Fig.~\ref{fig:g-2_Summary_Table}, the contributions to the anomalous muon magnetic moment for these points cover the range of $\Delta a_\mu \lesssim 20 \times 10^{-10}$, i.e.,  within 1$\sigma$ of the value of $\Delta a_\mu$ indicated by the data-driven theoretical estimate~\cite{Theory} and covering the ranges favored by lattice calculations~\cite{Lattice, Kuberski:2023qgx}, as well as the CMD-3 estimate~\cite{cmd} and the recent phenomenological analysis in~\cite{Davier:2023fpl}. The NUGM can provide values of $\Delta a_\mu$ significantly larger than those found in the CMSSM, constituting a major improvement over the CMSSM. 

As noted earlier, models with relatively large $\Delta a_\mu$ tend to have relatively low elastic scattering cross sections. The values for $\sigma_{\rm SI}$ and $\sigma_{\rm SD}$ for the three benchmark points are also given in Table \ref{table:bp1_d}. All three of these benchmark points have a Bino-like LSP whose relic density is determined primarily by stau coannihilation~\cite{stauco}.  The masses of the LSP, lighter chargino (Wino-like), left-handed smuon, lighter stau, gluino and lighter stop are also given in the Table. Their spectra are somewhat similar and have relatively low masses for the color-neutral states, which may be within reach of the LHC. However, the stop and gluino have in each case masses in excess of 10 TeV, beyond the reach of the LHC. The benchmarks are presented in Table \ref{table:bp1_d} and indicated by yellow symbols ($\bigtriangleup, \bigtriangledown$ and $\Diamond$)in the figures.   

We have also selected a set of three benchmark points for the NUHM3. In this case, even though there are many points for which the relic density is of order 0.12, these points are excluded since their spin-independent cross sections are too large, even when theoretical uncertainties are taken into account \cite{sospin }. We use Fig.~\ref{fig:NUHM3_gm2_planes} to select benchmark points that have a broad range of contributions to $\Delta a_\mu \lesssim 24 \times 10^{-10}$ but do not violate any of the constraints considered. {However, as discussed earlier, none of them saturate the Planck cold dark matter density.}
The benchmarks are presented in Table \ref{table:bp2} and indicated by yellow symbols ($\bigtriangleup, \bigtriangledown$ and $\Diamond$) in the figures. As in the case of the NUGM, we note that the masses of the color-neutral sparticles at these benchmarks are light enough to be potentially detectable at the LHC. Again as in the NUGM case, however, the colored sparticles are generally too heavy to be found at the LHC. An exception is the gluino at benchmark point A, which is lighter than the lower limit set in simplified models such as the CMSSM. However, as seen in the bottom panels of Fig.~2 of the second paper in~\cite{pmssm}, this limit may be relaxed in more general models, and the light gluino of point A may still be allowed and offer prospects for future detection at the LHC, a point requiring further study.

\clearpage

\begin{table}[h]
\begin{center}
$\begin{array}{ |c|c|c|c|c|c|c|c|   }
\hline
{} & M_{1/2} & m_{012}  & m_{03}  &  \mu  & M_A &  A_0/m_{012}  &  \tan\beta     \\
\hline
\hline
%  A  \quad ( \bigtriangleup )  &    570.69   &     439.46    &   8304.98   &    556.49     &     7339.81       &  18.27    &    24.75  \\
 A  \quad ( \bigtriangleup )  &    571   &     440    &   8305   &    557     &     7340       &  18.3    &    25  \\
%\hline
\hline
%B \quad ( \bigtriangledown )  &      1450.13       &      65.29     &     10374.00       &   142.92     &    9317.38       &     13.63     &  25.63 
B \quad ( \bigtriangledown )  &      1450       &      65      &     10374        &   143     &    9317        &     13.6     &  26   \\
\hline
%C  \quad ( \Diamond )  &  1237.55     &   837.16     &   11025.13        &     142.88      &     10238.91     &   16.40    &     29.61   \\
C  \quad ( \Diamond )  &  1238     &   837     &   11025         &     143      &     10239     &   16.4     &     30   \\
\hline
\end{array}$
\caption{\it Benchmark points for the NUHM3  model.}
\label{table:bp2}
\end{center}
\end{table}

\begin{table}[h]
\begin{center}
$\begin{array}{ |c|c|c|c|    }
\hline
{}  & A \quad ( \bigtriangleup )  &  B  \quad ( \bigtriangledown )  &  C  \quad ( \Diamond ) \\
\hline
\hline
\Omega_{\chi} h^2   & 3.24\times 10^{-2}      &  3.84\times 10^{-3}    &  4.11\times 10^{-3}     \\
\hline
\Delta a_{\mu} \, (10^{-10})  &    24.2      &   20.6       &   7.9       \\
\hline
\sigma_{SI} \, (pb)  &  2.45 \times 10^{-10}   &   9.99    \times 10^{-10}   &   1.23   \times 10^{-9}         \\
\hline
\sigma_{SI} \, (pb)\quad \text{re-scaled}  &  6.62 \times 10^{-11}   &   3.20    \times 10^{-11}   &  4.21   \times 10^{-11}         \\
\hline
\sigma_{SD} \, (pb)  &  9.53   \times 10^{-7}  &    4.51   \times 10^{-5}   &   6.15   \times 10^{-5}       \\
\hline
\sigma_{SD} \, (pb)\quad  \text{re-scaled} &  2.57   \times 10^{-7}  &    1.44    \times 10^{-6}   &    2.11   \times 10^{-6}       \\
\hline
\hline
m_\chi  &  247            &    153     & 153     \\
\hline
m_{\chi^+_1}  &  476           &   157  &  158          \\
\hline
m_{\tilde{\mu}_{L,R}}  &  261          &   205    &   719      \\
\hline
m_{\tilde{\tau}_1}  &  7759          &   9869    & 9791        \\
\hline
m_{\tilde{g}}  &  1388           &    3219   &   2829      \\
\hline
m_{\tilde{t_1}}  &  4837        &    7202    &  5785   \\
\hline
\end{array}$
\caption{\it Observables and masses for the NUHM3 benchmark points in Table~\ref{table:bp2}. 
At point A the LSP is a Bino, while at points B and C it is  a Higgsino.
The re-scaled direct detection cross sections ($\sigma \times\,\Omega_{\chi} h^2/0.12 $) for these points
are smaller than  the LZ bound, so these  points lie  below the black curve in Fig.~\ref{fig:NUHM3_gm2_planes}. }
\label{table:bp2_d}
\end{center}
\end{table}

%%%%%%%%%%%%%%%%%%%%%%%%%%%%%%%%%%%%%%%%%%%%%%%%
\section{Summary}
%%%%%%%%%%%%%%%%%%%%%%%%%%%%%%%%%%%%%%%%%%%%%%%%
\label{sec:concl}

We have shown in this paper that a significant supersymmetric contribution to the 
muon anomalous magnetic moment is possible if one relaxes the restrictive unification conditions on the gaugino and 
sfermion masses made in specific models such as the CMSSM and the NUHM2.  
In particular, in a  model with non-unified  gaugino masses one can find 
$\Delta a_\mu\sim 17 \times 10^{-10}$, e.g., if 
$M_{1,2} \sim 600\GeV $ and  $M_3\sim 8 \TeV$. In the case of non-universal Higgs and third-generation sfermion masses, one can find even larger $\Delta a_\mu \lesssim 24 \times 10^{-10}$ for first- and second-generation sfermion masses $ \sim 400  \GeV$ and 
third-generation sfermion masses  $ \sim 8  \TeV$. These possibilities are illustrated by the
benchmark points whose predictions for $\Delta a_\mu$ are shown in Fig.~\ref{fig:g-2_Summary_Table}.

\begin{figure}[ht!]
\centering
 \includegraphics[width=0.65\textwidth]{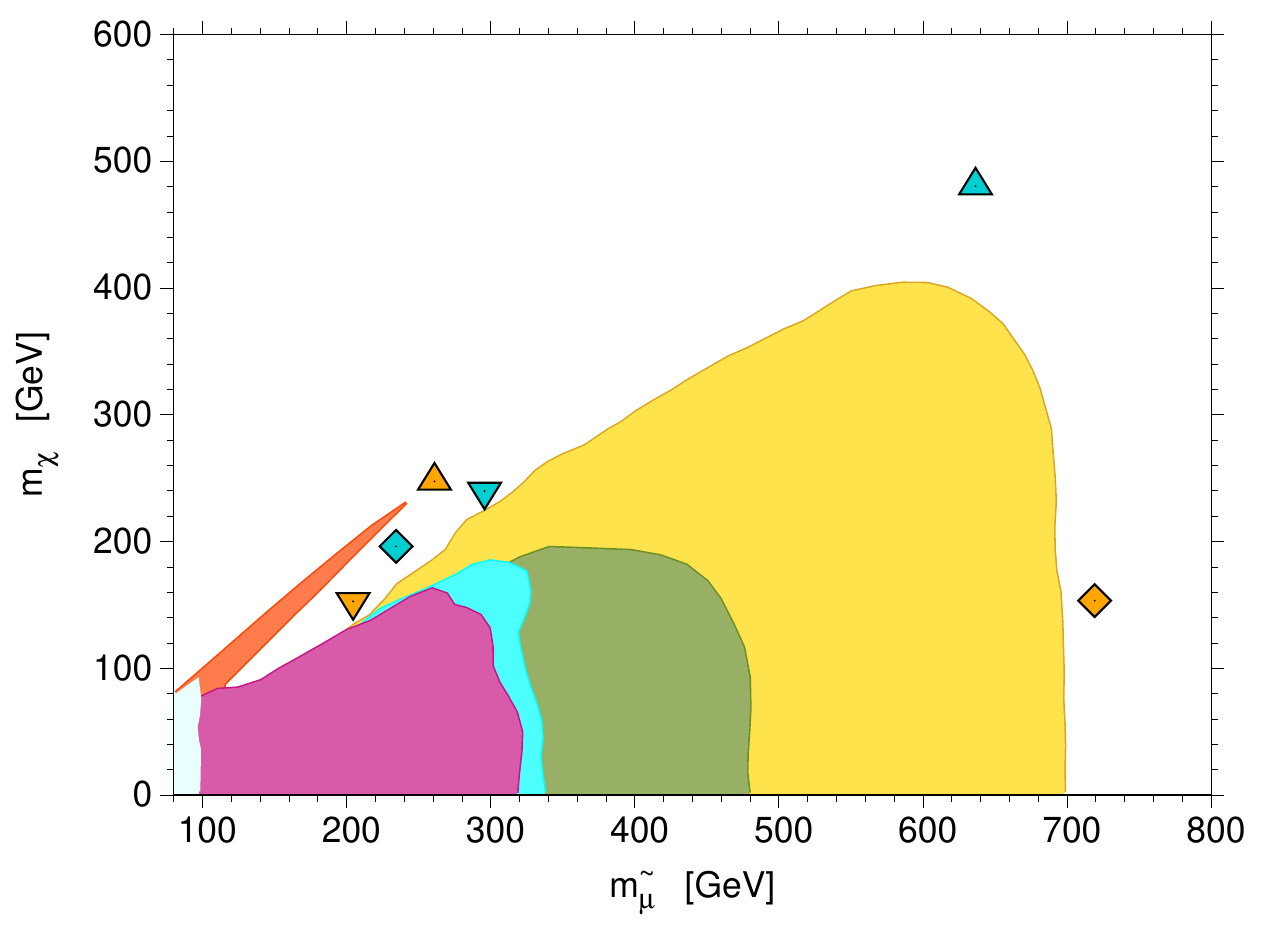}
 \caption{\it The benchmark  points for NUGM (cyan symbols) and NUHM3 (orange symbols), where triangle is point A,
inverted  triangle B and diamond C. The shaded regions are excluded by ATLAS dilepton searches at the LHC, see~\cite{slepton,Aad:2019qnd}: the colors correspond to different searches, as explained in~\cite{ATLAS:2023xco}.}
\label{fig:benchmarks_vs_LHC}
\end{figure}

These benchmark scenarios predict, in general, relatively light color-neutral sparticles that
may be detectable at the LHC. Fig.~\ref{fig:benchmarks_vs_LHC} shows the locations of the NUGM and NUHM3 benchmark points in the $(m_{\tilde \mu}, m_\chi)$ plane. We see that they are allowed by the current ATLAS constraints~\cite{LHCSUSY}, but likely to be vulnerable to foreseeable improvements in the LHC search sensitivity. On the other hand, the strongly-interacting sparticles must be heavier than the color-neutral sparticles,
in particular so as to yield a Higgs mass in the experimental range. A consequence of this
requirement is that the rates for direct detection of scattering on nuclei are 
typically small in the NUGM, lost in the `neutrino fog' except for some points with
$\Delta a_\mu < 5 \times 10^{-10}$. {However, the NUHM3 benchmark points 
have scattering rates that may be detectable after rescaling.}

%%%%%%%
%%%%%%%%%

We have shown in this paper how supersymmetric models could accommodate a discrepancy
$\Delta a_\mu$ between Standard Model prediction and the experimental measurements, 
whichever of the current theoretical estimates~\cite{Theory,Lattice,cmd,Davier:2023fpl} 
turns out to be more accurate. In all the benchmark scenarios studies, there are prospects
for detecting color-singlet sparticles at the LHC, and it is also possible that the gluino
might be within experimental reach. Many obituaries for supersymmetry have been pronounced,
but in the immortal words of Monty Python~\cite{MP}, it is ``not dead yet", and the anomalous magnetic moment of the muon may yet revive interest in supersymmetry.

%\newpage 
\section*{Acknowledgements}

 The work of J.E. was supported in part by the United Kingdom STFC Grant ST/T000759/1.
  The work of K.A.O. was supported in part by DOE grant DE-SC0011842 at the University of Minnesota.
V.C.S.  is grateful to the William I. Fine Theoretical Physics Institute 
at the University of Minnesota for their financial support 
and the warm hospitality extended to him  during
his sabbatical leave.

\end{document}